\documentclass[pre,twocolumn,preprintnumbers,amsmath,amssymb]{revtex4}

\usepackage{graphicx}% Include figure files
\usepackage{dcolumn}% Align table columns on decimal point
\usepackage{bm}% bold math
\usepackage{parskip}
\usepackage{color}
\usepackage{subfigure}
\usepackage{amssymb}
\usepackage{amsmath}
\usepackage{amssymb}
\usepackage{graphicx}

\def\be{\begin{eqnarray}}
\def\ee{\end{eqnarray}}
\def\be{\begin{equation}}
\def\ee{\end{equation}}
\begin{document}

\title{Multiparameter universality and intrinsic diversity \\ in weakly anisotropic bulk and confined systems  }

\author{Volker Dohm}

\affiliation{Institute for Theoretical Physics, RWTH Aachen University, 52056 Aachen, Germany}

\date {November 5, 2023}

\begin{abstract}
An overview of recent advances in the theory of critical phenomena in $d$-dimensional weakly anisotropic systems is given. On the basis of a generalized shear transformation between anisotropic and isotropic systems, exact and approximate results are discussed for bulk and confined systems in two and three dimensions where conformal field theory
and the minimal renormalization without $\varepsilon$-expansion
play a crucial role. Stimulation for this research comes from the seminal work  by V. Privman and M.E. Fisher in 1984 in which the principle of two-scale-factor universality for bulk systems has been extended to finite systems. Based on this principle for isotropic systems we predict the validity of multiparameter universality with up to $d(d+1)/2+1$ nonuniversal parameters in $d$-dimensional anisotropic bulk and confined systems with periodic boundary conditions (BC). The verification of multiparameter universality for confined anisotropic systems with realistic BC and the study of the intrinsic diversity of the critical behavior of magnetic materials, superconductors, liquid crystals, and solids with non-cubic symmetry
are a major challenge to future  research.
\end{abstract}
\maketitle

\renewcommand{\thesection}{\Roman{section}}
\renewcommand{\theequation}{1.\arabic{equation}}
\setcounter{equation}{0}
\section{Introduction}
The traditional theory of ordinary critical phenomena with short-range interactions \cite{fish-1,bre-1,priv,pri,hohenberg1976,cardybuch,pelissetto,zinn2007} including conformal field theory \cite{franc1997,cardy1987} as well as the more recent development of the functional renormalization group \cite{metzner2021} have been primarily focussed on the critical behavior in {\it spatially isotropic} systems, apart from anisotropic two-dimensional models \cite{Vaidya1976,night1983,WuCoy,CoyWu,kim1987,Perk1,Perk2,Perk3,Perk4,NO1999,HH2019,aharony1980,Izmailian}. Spatial anisotropy is a fundamental property that is omnipresent in condensed matter physics where it is the origin of a wide variety of nonuniversal effects. Substantial evidence has emerged over the last two decades \cite{cd2004,dohm2005,selke2005,selke2009,dohm2006,chen-zhang,dohm2008,dohmphysik2009,DG,dohm2011,kastening-dohm,dohm2018,dohm2019,DW2021,DWKS2021} that part of this nonuniversal diversity persists even in the critical region as well as in the near-critical Goldstone regime  of so-called weakly anisotropic systems. This includes magnetic materials, superconductors, alloys, liquid crystals, compressible solids and solids with structural phase transitions. Recently an unexpected complex form of self-similarity of anisotropy effects in finite weakly anisotropic systems with periodic boundary conditions (BC) has been found \cite{DW2021} near the instability where weak anisotropy breaks down.

Ordinary bulk critical phenomena can be divided into universality classes characterized
by the dimension $d$
and the symmetry properties of the ordered state \cite{fish-1,priv,pelissetto}.
As an example we consider $O(n)$-symmetric systems with short-range interactions and
with an $n$-component order parameter.
Within each $(d,n)$ universality class, all systems have the same universal quantities
(critical exponents, amplitude ratios, and scaling functions) which includes isotropic and weakly anisotropic systems since spatial anisotropy is only a marginal perturbation in the renormalization-group sense \cite{pri,priv,Aharony1976,weg-1,bruce,zia}. The principle of two-scale-factor universality
(or hyperuniversality) \cite{stau,priv,aha-74,hohenberg1976,pri,weg-1,gerber,privmanbuch}
predicts that, once the universal quantities of a universality class are known,
the asymptotic critical behavior of any particular system of this universality class is known completely provided that only two nonuniversal amplitudes are given.
This principle was stated to be valid for all systems in a universality class
\cite{priv,henkelprivmanbuch,privman1988}. Furthermore it was asserted \cite{cardy1987,cardy1983,cardybuch,Indekeu,binder-wang,nightingaleprivmanbuch,barber1984,zia} that asymptotic isotropy can be restored in weakly anisotropic systems by a suitable anisotropic scale transformation and that universality can be restored \cite{Indekeu}, reintroduced \cite{nightingaleprivmanbuch}, or repaired \cite{priv} in some cases. However the question was left unanswered how the various nonuniversal effects due to non-cubic anisotropy in bulk \cite{bruce,cardybuch,Aharony1976,aharony1980,Vaidya1976,WuCoy,Perk1,Perk2,Perk3,Perk4} and confined \cite{cardyfinite,cardy1987,kim1987,Yurishchev,Indekeu,night1983,nightingaleprivmanbuch,NO1999,barber1984,binder-wang} systems could be reconciled with the principle of two-scale-factor universality.

A systematic study of the effect of non-cubic spatial anisotropy on the critical behavior in bulk and confined systems was begun \cite{cd2004} by introducing a nondiagonal anisotropy matrix ${\bf A}$ into the $\varphi^4$ field theory. An anisotropy-induced nonunversality was discovered for the critical Casimir amplitude \cite{DDPhysRep,bloete,krech} and critical Binder cumulant ratio \cite{priv} which were the hallmarks of finite-size universality. These results demonstrated that restoring isotropy relates the critical behavior of anisotropic systems to that of isotropic systems but does not restore  two-scale-factor universality. Shortly thereafter the nonuniversality of the Binder cumulant ratio was confirmed by Monte Carlo simulations of an anisotropic Ising model \cite{selke2005}. The analytic results \cite{cd2004} were based on the exact large-$n$ limit and on the $\varphi^4$ theory in the minimal subtraction scheme in three dimensions \cite{dohm1985,schl,schl1990}. The violation of two-scale-factor universality in weakly anisotropic bulk and confined systems was subsequently confirmed \cite{dohm2006,DG,dohm2008,dohm2018,dohm2019,chen-zhang,selke2009,kastening-dohm}, most recently by exact analytic results for the critical free energy and Casimir amplitude in two dimensions \cite{DW2021} based on conformal field theory \cite{franc1997}, as well as by Monte Carlo simulations \cite{DWKS2021}.

Weakly anisotropic systems do not have a unique single
bulk correlation length but rather an angular-dependent correlation length \cite{dohm2019} with $d$ independent nonuniversal amplitudes in the $d$ principal directions
which are determined by $d(d-1)/2$ nonuniversal angles. This has led to the hypothesis \cite{dohm2018} (first formulated \cite{dohm2008} for bulk amplitude relations) that in weakly anisotropic bulk and confined systems two-scale-factor universality is absent and is replaced by  multiparameter universality which allows for up to $d(d+1)/2+1$ independent nonuniversal parameters. This implies a revised notion of a universality class: it must be divided into subclasses \cite{dohm2008,dohmphysik2009} of isotropic and weakly anisotropic systems with different scaling forms for the  bulk correlation function and for the free energy of confined systems within the same universality class. These anisotropic scaling forms are governed by a reduced anisotropy matrix ${\bf \bar A}$ which has a universal structure in terms of principal correlation lengths and principal directions where the latter depend in a generically unknown way on the anisotropic interactions. So far there exists no general proof for this hypothesis. In the following we report on recent advances \cite{dohm2018,dohm2019,DW2021,dohm2023,DKW2023,KWD2023} of such a proof based on the validity of two-scale-factor universality in isotropic bulk \cite{hohenberg1976,priv,weg-1,stau} and confined \cite{pri,priv} systems and on a generalized shear transformation \cite{dohm2023} between anisotropic and isotropic systems. Among the $d(d+1)/2+1$ nonuniversal parameters of weakly anisotropic systems there are only two parameters that can be determined by macroscopic thermodynamic measurements \cite{dohm2023}
whereas the $d(d+1)/2-1$ independent parameters contained in  ${\bf \bar A}$ cause a high degree of intrinsic diversity arising from
the nonuniversal angular dependence of the critical correlations.
\renewcommand{\thesection}{\Roman{section}}
\renewcommand{\theequation}{2.\arabic{equation}}
\setcounter{equation}{0}
\section{${\bf O(n)}$-symmetric models with lattice anisotropy}
In order to discuss the issue of multiparameter universality we analyze different types of $O(n)$-symmetric anisotropic models that belong to the same universality classes. As examples we consider (i) the $\varphi^4$ model and (ii) the $n$-vector model.

(i) The $\varphi^4$ lattice Hamiltonian  and  the total free energy divided by $k_BT$ are
\begin{eqnarray}
\label{2a} H  &=&   v \Bigg[\sum_{i=1}^N \left(\frac{r_0}{2}
\varphi_i^2 + u_0 (\varphi_i^2)^2 \right)
\nonumber\\
&+& \sum_{i, j=1}^N \frac{K_{i,j}} {2} (\varphi_i -
\varphi_j)^2 \Bigg],\\ \;\;\;\;\;\;\;\;
\label{Ftotphi}
{\cal F}_{\rm tot}&=&-\ln\Big[\prod_{i = 1}^{ N} \frac{\int
d^n {\varphi}_i}{v^{n (2-d) / (2d)}} \Big] \exp \left(- H \right).
\end{eqnarray}
The variables $\varphi_i \equiv \varphi ({\bf x}_i)$ are $n$-component vectors
on $N$ lattice points ${\bf x}_i \equiv(x_{i1}, x_{i2},\ldots, x_{id})$ of a
$d$-dimensional Bravais lattice of volume $V = Nv$
where $v$ is the volume of the primitive cell.
We assume periodic BC.
The large-distance anisotropy arising from the couplings $K_{i,j}$ is described by a dimensionless symmetric anisotropy matrix ${\bf A}$ \cite{cd2004} with matrix elements \cite{dohm2006}
\begin{equation}
\label{2i}
A_{\alpha \beta} =  N^{-1} \sum^{ N}_{i,
j = 1} (x_{i \alpha} - x_{j \alpha}) (x_{i \beta} - x_{j \beta})\;K_{i,j}.
\end{equation}
The continuum version of this model in terms of the vector
field $\varphi({\bf x})$ has the Hamiltonian \cite{cd2004,dohm2008}
\begin{eqnarray}
\label{contin}
&&H_{\text {field}} = \nonumber\\
&&\int_V d^d x \Big[\frac{r_0}{2}
\varphi^2 + \sum_{\alpha,
\beta=1}^d \frac{A_{ \alpha \beta}}{2} \frac{\partial \varphi}
{\partial x_\alpha} \frac{\partial \varphi} {\partial x_\beta}
 + u_0 (\varphi^2)^2  \Big].\;\;\;\;
\end{eqnarray}
Weakly anisotropy systems have the same critical exponents as the isotropic system which requires \cite{cd2004}
$\det {\bf A}(\{K_{i,j}\})>0$.
We shall discuss the exact structure of the anisotropic bulk correlation function
\begin{eqnarray}
\label{corrfctphi}
 G({\bf x_i}-{\bf x_j},t)
= \lim_{V \to \infty}\big[<\varphi ({\bf x}_i) \cdot \varphi ({\bf x}_j)> - <{\varphi}> ^2\big]\;\;\;\;\;
\end{eqnarray}
in $d\geq 2$ dimensions. Near $T_c$ the decomposition
\begin{eqnarray}
\label{singnonsing}
{\cal F}_{\rm tot} ={\cal F}_s + {\cal F}_{ns}
\end{eqnarray}
into singular and nonsingular parts is appropriate.
We shall discuss the singular part $f_s={\cal F}_s/V $ of the free energy density $f={\cal F}_{\rm tot}/V$  in a $d=3$ block geometry \cite{dohm2018} as well as the excess free energy
\begin{eqnarray}
\label{FtotphiMEF}
{\cal F}^{\rm ex} ={\cal F}_{\rm tot} - {\cal F}_b
\end{eqnarray}
and the critical free energy of the finite system at $T_c$
\begin{eqnarray}
\label{FcritTc}
{\cal F}_c = \lim_{T\to T_c}{\cal F}^{\rm ex}  = \lim_{T\to T_c}{\cal F}_s
\end{eqnarray}
in rectangular and parallelogram geometries \cite{DW2021,dohm2023,DKW2023} where ${\cal F}_b = V f_b$ is the bulk part of ${\cal F}_{\rm tot}$ with $f_b=\lim_{V\to \infty} {\cal F}_{\rm tot}/V$ being the bulk free-energy density.

(ii) The anisotropic $O(n)$-symmetric $n$-vector model is defined on the same lattice with the same BC as the $\varphi^4$ model. It has the Hamiltonian and total free energy (divided by $k_B T$)
\begin{eqnarray}
\label{Hspin}
H^{\rm sp}& =& - \sum_{i,j} E_{i,j} S_i \cdot S_j,\\
{\cal F}^{\rm sp}_{\text{tot}}&=&-\ln\Big[\prod_{i = 1}^{N} \int
d^n S_i \Big] \exp \left(- \beta H^{\rm sp}\right)
\end{eqnarray}
with pair interactions $E_{i,j}$ and $\beta =1/(k_B T)$. The continuous spin variables $S_i$ are $n$-component vectors with a fixed length  $S_i^2=1$. The definitions of $G^{\rm sp},{\cal F}^{\rm sp}_s,
{\cal F}^{\rm sp,ex}$, and  ${\cal F}^{\rm sp}_c$ are analogous to (\ref{corrfctphi})-(\ref{FcritTc}). We shall discuss the exact structure of $G^{\rm sp}$ for $d\geq 2$ and exact results for
${\cal F}^{\rm sp,ex}$ and ${\cal F}^{\rm sp}_c$ in $d=2$  dimensions.

For $n=1,2,3, \infty$ the models (i) and (ii) belong to the Ising, $XY$, Heisenberg and spherical universality classes.  In forthcoming papers \cite{dohm2023,DKW2023,KWD2023} it is shown that the hypothesis of multiparameter universality \cite{dohm2018} is indeed valid for  $G,G^{\rm sp}$ in $d\geq 2$ dimensions for general $n$ and for ${\cal F}^{\rm ex},{\cal F}^{\rm sp, ex}$  in the $(d=2,n=1)$ Ising universality class.
Our strategy is to employ exact or approximate results for $G^{\rm iso},G^{\rm sp,iso}$ and ${\cal F}^{\rm ex,iso},{\cal F}^{\rm sp, ex,iso}$ of isotropic systems for which two-scale-factor universality \cite{hohenberg1976,pri,priv,weg-1} can be invoked, and to derive the structure of $G,G^{\rm sp}$ and ${\cal F}^{\rm ex},{\cal F}^{\rm sp, ex}$ of weakly anisotropic systems by means of nonuniversal inverse shear transformations \cite{dohm2023}.

An extension of this analysis to the Gaussian universality class is given in Refs. [63] and [64].
\renewcommand{\thesection}{\Roman{section}}
\renewcommand{\theequation}{3.\arabic{equation}}
\setcounter{equation}{0}
\section{Isotropic case: Two-scale-factor universality}
The couplings $K_{i,j}$ and $E_{i,j}$ can be chosen such that the bulk system has isotropic correlations in the large-distance scaling regime near $T_c$. The isotropic  bulk correlation function has  the established scaling form for general $n$ above and at $T_c$ $(+)$ and $n=1$ below $T_c$ $(-)$
\cite{pri,dohm2019,dohm2023}
\begin{eqnarray}
\label{3c} G^{\rm iso} (|{\bf x}|, t) &=& \frac{\Gamma_+^{\rm iso}(\xi^{\rm iso}_{0+})^{-2+\eta}}{ | {\bf x} |^{ d - 2 +\eta}}
\;\Psi_\pm \Big(\frac{|{\bf x}|}{ \xi^{\rm iso}_{\pm}(t)}\Big) \; ,\;\;\;\;\;\;
\\
\label{isocorr}
\xi^{\rm iso}_\pm(t)&=&\xi^{\rm iso}_{0\pm}\;|t|^{-\nu}, \;\; t=(T-T_c)/T_c
\end{eqnarray}
with the universal critical exponents $\eta$ and $\nu$, the universal scaling function $\Psi_\pm$, and the nonuniversal amplitudes $\Gamma^{\rm iso}_\pm$ and $\xi^{\rm iso}_{0\pm}$ of the susceptibility $\chi^{\rm iso}(t)=\Gamma^{\rm iso}_\pm |t|^{-\gamma}$ and of the correlation lengths, respectively.
 The latter are universally related as
\begin{eqnarray}
\label{ratioxi}
\xi^{\rm iso}_{0 +}/\xi^{\rm iso}_{0 -}=X_\xi= {\rm universal}.
\end{eqnarray}
The scaling function $\Psi_\pm$ depends on the universality class. It is known exactly for the $d=2$ Ising universality class \cite{dohm2019}, for the spherical universality class corresponding to the large-$n$ limit \cite{dohm2023}, and for the Gaussian universality class corresponding to the $\varphi^4$ Hamiltonians (\ref{2a}) and  (\ref{contin}) with $u_0=0$ \cite{dohm2023}. We recall that the universal scaling regime is defined by the limit $| {\bf x} | \to \infty, \xi^{\rm iso}_{\pm}(t)\to \infty $ at fixed finite ratio $ | {\bf x} |/ \xi^{\rm iso}_{\pm}(t)$ whereas  scaling and universality are not valid in the regime $| {\bf x} | \gg \tilde a, \xi^{\rm iso}_{\pm}(t) \gg \tilde a, | {\bf x} | \gg \xi^{\rm iso}_{\pm}(t)$ even arbitrarily close to $T_c$ where $\tilde a$ is some microscopic reference length \cite{dohm2008}.

The singular part of the bulk free-energy density $f^{\text {iso}}_{b,s,\pm}$ of isotropic systems has the asymptotic form \cite{priv,pelissetto,dohm2023}
\begin{eqnarray}
\label{3a}
f^{\text {iso}}_{b,s,\pm} (t) =\left\{
\begin{array}{r@{\quad \quad}l}
                         \; A^{\text {iso}}_\pm |t|^{d \nu}\quad          & \mbox{for} \;2<d<4\;, \\
                         \; \frac{1}{2}A^{\rm iso}_\pm |t|^{2\nu}\ln |t| & \mbox{for} \;d=2 \;
                \end{array} \right.
\end{eqnarray}
where the nonuniversal amplitudes $A^{\rm iso}_\pm $ have the universal ratio
$A^{\rm iso}_+/ A^{\rm iso}_-$. According to two-scale-factor universality for isotropic bulk systems \cite{hohenberg1976,priv,stau,weg-1}
these amplitudes
are universally related to the correlation-length amplitude $ \xi_{0+}^{\rm iso}$  through
\begin{eqnarray}
\label{3fy}
 \left(\xi_{0+}^{\rm iso}\right)^d A_+^{\rm iso} = \; Q_1
\end{eqnarray}
with  the universal constant $Q_1$. The amplitude $A_+^{\rm iso}$ enters the singular part of the specific heat $C^{\rm iso}_{b,s,+}(t)=-\partial^2f^{\rm iso}_{b,s,+}(t)/\partial t^2$, thus the correlation length $\xi_{0+}^{\rm iso}$ is determined by the thermodynamic amplitude of the specific heat \cite{dohm2023}. The important consequence of two-scale-factor universality for isotropic bulk systems is that all nonuniversal parameters of the isotropic bulk correlation function can be completely determined by the thermodynamic amplitudes of the bulk susceptibility and the bulk specific heat without involvement of measurements of the spatial dependence of the correlation function via scattering experiments. We shall show in Sec. IV that weak anisotropy destroys this important feature due to the intrinsic nonuniversality of the anisotropic bulk correlation function.

The hypothesis of two-scale-factor universality for {\it finite} systems (with a large characteristic length $L_0$, a small ordering field $h$, and a small reduced temperature $|t|$) in a $L_1 \times L_2 \cdot \cdot \cdot \times L_d $ block geometry with the volume $V$ has been  formulated in terms of the singular part of the free-energy density in the Privman-Fisher scaling form as \cite{pri}
\begin{eqnarray}
\label{freesing}
&&f_s(t,h,L_1,L_2,...) \nonumber\\
&&= L_0^{-d} \; Y(C_1 t L_0^{1/\nu}, C_2 h L_0^{\Delta/\nu}; L_1/L_0, L_2/L_0,...),\;\;\;\;\;\;
\end{eqnarray}
with universal critical exponents $\nu,\Delta$ and, for given BC,  with the universal scaling function $Y$. The shape ratios $L_j/L_0$ are supposed to approach finite constants as $L_0 \to \infty$. The two metric factors $C_1$ and $C_2$ are the only nonuniversal parameters entering (\ref{freesing}). With the choice $L_0=V^{1/d}$ we see that the Privman-Fisher scaling function $Y$ has the meaning of the singular part of the free energy
\begin{eqnarray}
\label{freesingenergy}
&&{\cal F}_s=Vf_s \nonumber \\
&&=  Y(C_1 t V^{1/(d\nu)}, C_2 h V^{\Delta/(d\nu)}; L_1/V^{1/d}, L_2/V^{1/d},...).\;\;\;\;\;\;\nonumber\\
\end{eqnarray}
 In particular we shall consider the critical free energy
\begin{eqnarray}
\label{criticalF}
{\cal F}_c=\lim_{t \to 0,h\to 0}{\cal F}_s=
% = F^{\rm ex,iso}\big(0,\alpha, \rho_{\rm p}^{\rm iso}\big)
%=\lim_{T\to T_c}{\cal F}_s
 Y(0,0; L_1/V^{1/d}, L_2/V^{1/d},...).
\end{eqnarray}
The constants $C_1$ and $C_2$ are universally related to the
constants $A_1$ and $A_2$ of  the singular part of the bulk free-energy density \cite{pri}
\begin{equation}
\label{1abulk} f_{b, s} (t, h)=\lim_{\{L_j\}\to \infty}f_s = A_1 |t|^{d \nu} \; W_\pm (A_2 h
|t|^{-
 \Delta})
\end{equation}
which can be measured via the bulk specific heat and susceptibility.
Because of the logarithmic factor in (\ref{3a}) for $d=2$ it is necessary to decompose $f_s =f_{b,s} +f^{\rm ex}_s $ and to reformulate the scaling forms (\ref{freesing}) and (\ref{freesingenergy}) for $d=2$ only for $f^{\rm ex}_s $ and ${\cal F}^{\rm ex}_s $ (where we have omitted a possible surface contribution for nonperiodic BC).

The function $Y$ has been asserted \cite{priv,henkelprivmanbuch,privman1988} to be the same for all systems in a given universality class in $d<4$ dimensions.
While the validity of the  structure of (\ref{3c})-(\ref{criticalF}) is well established for isotropic systems, the hypothesis of multiparameter universality \cite{dohm2018} predicts weak anisotropy to cause significant changes of these structures. This is confirmed by recent advances \cite{dohm2019,dohm2023,DW2021,DWKS2021,DKW2023,KWD2023} to be discussed below.
\renewcommand{\thesection}{\Roman{section}}
\renewcommand{\theequation}{4.\arabic{equation}}
\setcounter{equation}{0}
\section{Multiparameter universality and intrinsic diversity of the bulk correlation function}
In this section we present the exact scaling forms of the anisotropic bulk correlation functions of both the $\varphi^4$ and the $n$-vector models in $d\geq 2$ dimensions and show the validity of multiparameter universality.
\subsection{Anisotropic bulk correlation function of the $\varphi^4$ theory}
We first sketch the derivation of our result \cite{dohm2018} for $G({\bf x},t)$ in the anisotropic $\varphi^4$ theory. It is possible to relate $G({\bf x},t)$ to the correlation function $G'(|{\bf x'}|,t)$ of the isotropic system. For this purpose a shear transformation ${\bf x} \to {\bf x'}$ is performed by \cite{cd2004,dohm2023}
\begin{eqnarray}
\label{shearalt}
{\bf x'}&=& {\mbox {\boldmath$\lambda$}}^{-1/2} {\bf U}{\bf x}, \\
\label{shearphi}
\varphi'({\bf x'})&=&(\det {\bf A})^{1/4}\varphi({\bf x}),\\
\label{shearu}
u'_0&=&(\det{\bf A}) ^{-1/2}u_0
\end{eqnarray}
where ${\bf A}$ is given by (\ref{2i}).
It consists of a rotation
provided by the orthogonal matrix
${\bf U}$ and a subsequent spatial rescaling by the diagonal rescaling matrix
${\mbox {\boldmath$\lambda$}}$ such that
the Hamiltonian (\ref{contin}) is brought into the isotropic form
\begin{eqnarray}
\label{2zxx}
H_{\text {field}}&=& H_{\text {field}}'\\
\label{isotrH}
&=& \int_{V'} d^d x'
\Big[\frac{r_0} {2} \varphi'({\bf x}')^2 +  \frac{1} {2} (\nabla'
\varphi')^2 +
 u'_0 (\varphi'^2)^2\Big]\;\;\;\;\;\;\;\;
\end{eqnarray}
with the asymptotic bulk correlation function $G'(|{\bf x'}|,t)$ for $V' \to \infty$.
The matrix elements of ${\bf U}(\{{\bf e}^{(\alpha)}\})$ are $U_{\alpha \beta} = e_{\beta}^{(\alpha)}$
which are determined by the $d$ eigenvectors ${\bf e}^{(\alpha)}$ of the matrix ${\bf A}$
whose Cartesian components are denoted by $e_{\beta}^{(\alpha)}$.
They satisfy the eigenvalue equation
${\bf A}
\;{\bf e}^{(\alpha)}={\mbox {$\lambda$}_{ \alpha}}\;{\bf e}^{(\alpha)},\;\alpha = 1, 2, ..., d$
with ${\bf e}^{(\alpha)}\cdot{\bf e}^{(\beta)}=\delta_{\alpha\beta}$. The matrix ${\bf U}$ diagonalizes the matrix ${\bf A}$ according to
\label{lambdaUAU}
\begin{eqnarray}
{\mbox {\boldmath$\lambda$}}={\bf U}{\bf A}{\bf U}^{-1}
\end{eqnarray}
whose diagonal elements are the eigenvalues $\lambda_\alpha>0$ of ${\bf A}$.
The eigenvectors ${\bf e}^{(\alpha)}$ determine the directions of the principal axes of the large-distance correlations of the anisotropic system
\cite{cd2004,dohm2008,dohm2018}.
Because of the transformation of $\varphi$, (\ref{shearphi}), the anisotropic bulk correlation function (\ref{corrfctphi})
is transformed as
\begin{eqnarray}
\label{isocorrf}
G'(|{\bf x'}|,t)&=& (\det {\bf A})^{1/2} G({\bf x},t).\;\;\;\;\;
\end{eqnarray}
The bulk susceptibility  $\chi(t)=\Gamma_\pm |t|^{-\gamma}=\chi'(t)=\Gamma'_\pm |t|^{-\gamma}$ remains invariant, i.e., $\Gamma'_\pm =\Gamma_\pm $, as follows from the sum rules \cite{dohm2008,dohm2023}
\begin{eqnarray}
\label{relationchiphi}
&&\chi'(t)=\int d^d{\bf x'}\;\;G' (|{\bf x'}|, t)
= \int d^d{\bf x}\;\;G ({\bf x}, t)=\chi(t).\;\;\;\;\;\;\;
\end{eqnarray}
From the isotropic scaling form (\ref{3c}) we have
\begin{eqnarray}
\label{3cprime} G' (|{\bf x'}|, t) &=& \frac{\Gamma'_+(\xi'_{0+})^{-2+\eta}}{ | {\bf x'} |^{ d - 2 +\eta}}
\;\Psi_\pm \Big(\frac{|{\bf x'}|}{ \xi'_{\pm}(t)}\Big) \; ,\;\;\;\;\;\;
\\
\label{isocorrprime}
\xi'_\pm(t)&=&\xi'_{0\pm}\;|t|^{-\nu},
\end{eqnarray}
where $\xi'_\pm(t)$ is the correlation length of the isotropic system (\ref{2zxx}). The amplitudes $\xi'_{0+}$ and  $\xi'_{0-}$ are universally related by
\begin{eqnarray}
\label{ratioampprime}
\xi'_{0+}/\xi'_{0-}=X_\xi
\end{eqnarray}
where $X_\xi$ is the same universal constant as in (\ref{ratioxi}). The principal correlation lengths of the anisotropic system above and below $T_c$ have been determined as \cite{cd2004,dohm2008}
\begin{eqnarray}
\label{xialphaxz}
\xi^{(\alpha)}_{\pm}(t)&=&\xi^{(\alpha)}_{0\pm}|t|^{-\nu}={\mbox {$\lambda$}_{\alpha}}^{1/2}\xi'_\pm(t).
\end{eqnarray}
As a consequence their amplitude ratios are independent of the direction $\alpha$ and are universal quantities given by
\begin{eqnarray}
\label{ratioamp}
\xi^{(\alpha)}_{0+}/\xi^{(\alpha)}_{0-}=\xi'_{0+}/\xi'_{0-}=X_\xi \;\; \text {for each}\;\; \alpha
\end{eqnarray}
where $X_\xi$ is independent of $\alpha$ and the same universal constant as in (\ref{ratioxi}).

It has been shown recently \cite{dohm2018,dohm2023} that by inverting the shear transformation one obtains from  $G({\bf x},t)=(\det {\bf A})^{-1/2}G'(|{\bf x'}|,t) $ the correlation function of the anisotropic $\varphi^4$ model for general $n\geq 1, T\geq T_c$ and  for $n=1, T<T_c$ in $2\leq d <4$ dimensions
\begin{eqnarray}
\label{Gphi4aniso}
 &&G({\bf x},t)=\frac{\Gamma_+(\bar \xi_{0+})^{-2+\eta}}{ [{\bf x}\cdot {\bf \bar A}^{-1}{\bf x}]^{(d-2+\eta)/2}} \Psi_\pm \Big(\frac{[{\bf x}\cdot {\bf \bar A}^{-1}{\bf x}]^{1/2}}{\bar \xi_\pm(t)}\Big)\;\;\;\;\;\;\;\;\;
 \end{eqnarray}
with the same universal scaling function $\Psi_\pm $ as for the isotropic system where
\begin{eqnarray}
\label{ximeanx}
\bar \xi_{\pm}(t)&=&\big[\prod^d_{\alpha = 1} \xi_{\pm}^{(\alpha)}(t)\big]^{1/d}=\bar \xi_{0\pm}|t|^{-\nu},\\
 \bar \xi_{0\pm}&=&\big[\prod^d_{\alpha = 1} \xi_{0\pm}^{(\alpha)}\big]^{1/d}
\end{eqnarray}
is the mean correlation length and
 \begin{eqnarray}
\label{Aquerspecial}
  {\bf \bar A}&=&{\bf A}/(\det {\bf A})^{1/d}={\bf U}^{-1}{\bar{\mbox {\boldmath$ \lambda$}}}{\bf U}
 \end{eqnarray}
is the reduced anisotropy matrix with
the diagonal reduced rescaling matrix
\begin{eqnarray}
\label{lambdaquerspecial}
 {\bar{\mbox {\boldmath$ \lambda$}}}= {{\mbox {\boldmath$\lambda$}}}  /(\det {\mbox {\boldmath$\lambda$}})^{1/d}.
 \end{eqnarray}
Its diagonal elements are expressed in terms of ratios of principal correlation lengths as
\begin{eqnarray}
\label{lambdaalphaxixz}
\bar \lambda_\alpha\big(\{\xi_{0 \pm}^{(\alpha)}\}\big)&=& \prod^d_{\beta=1,\; \beta \neq\alpha}\Big(\frac{\xi_{0\pm}^{(\alpha)}}{\xi_{0\pm}^{(\beta)}}\Big)^{2/d}
=\Big(\frac{\xi_{0\pm}^{(\alpha)}}{\bar \xi_{0\pm}}\Big)^2.
\end{eqnarray}
Here the isotropic correlation length $\xi'_{0\pm}$ has been canceled.
Thus, instead of expressing ${\bf \bar A}$ as a nonuniversal function of the couplings $K_{i,j}$ we obtain the structure
\begin{eqnarray}
\label{barAxyy}
{\bf \bar A} ={\bf \bar A}\big(\{\xi_{0\pm}^{(\alpha)},{\bf e}^{(\alpha)}\}\big)={\bf U(\{{\bf e}^{(\alpha)}\})}^{-1}{\bf \bar{\mbox {\boldmath$\lambda$}}} \big(\{\xi_{0 \pm}^{(\alpha)}\}\big) {\bf U(\{{\bf e}^{(\alpha)}\})}\nonumber\\
\end{eqnarray}
in terms of ratios of principal correlation lengths. The appearance of the matrix  ${\bf \bar A}$ in (\ref{Gphi4aniso}) violates two-scale-factor universality and introduces an intrinsic diversity of the correlation function.
\subsection{Anisotropic bulk correlation function of the $n$-vector model}
As a representative of weakly anisotropic systems other than the $\varphi^4$ model we take the fixed-length $n$-vector Hamiltonian $H^{\rm sp}$, (\ref{Hspin}), for general $d$ and $n$. The special shear transformation (\ref{shearalt})-(\ref{shearu}) of the $\varphi^4$ theory
is not applicable to  the $n$-vector model (\ref{Hspin}) where it is unknown how to construct an appropriate continuum Hamiltonian with a large-distance anisotropy matrix as a function of the couplings $E_{i,j}$ that plays the same role as ${\bf A}$ in the anisotropic $\varphi^4$ theory. Instead we have introduced \cite{dohm2023} a generalized shear transformation that does not need an anisotropy matrix corresponding to ${\bf A}$ and that is applicable to all weakly anisotropic systems. Our only assumptions for the  $n$-vector model (\ref{Hspin}) are
(i) that there exist $d$ principal unit vectors ${\bf e}^{{\rm sp}(\alpha)}$  in the directions of $d$  principal axes together with $d$ principal correlation lengths above and below $T_c$
\begin{eqnarray}
\label{princcorr}
\xi^{{\rm sp}(\alpha)}_{\pm}(t)=\xi^{{\rm sp}(\alpha)}_{0\pm}|t|^{-\nu}
\end{eqnarray}
along these directions where $\nu$ is the critical exponent of the isotropic $n$-vector model, and (ii) that the amplitude ratios  satisfy
\begin{eqnarray}
\label{ratioalpha}
\xi^{{\rm sp}(\alpha)}_{0+}/\xi^{{\rm sp}(\alpha)}_{0-}= X_\xi \;\;\text {for each}\;\; \alpha
\end{eqnarray}
where $X_\xi$ is the same universal constant as in (\ref{ratioxi}) for isotropic systems of the $(d,n)$ universality class.

We introduce the orthogonal set of $d$ {\it principal correlation vectors}
\begin{eqnarray}
\label{corrvectorsp}
{\mbox {\boldmath$\xi$}}_{0\pm}^{{\rm sp}(\alpha)}=\xi^{{\rm sp}(\alpha)}_{0\pm}{\bf e}^{{\rm sp}(\alpha)}, \;\;\; \alpha=1,...,d.
\end{eqnarray}
We use a Cartesian coordinate system with orthogonal unit vectors ${\mbox {\boldmath$\epsilon$}}^{(\alpha)}$, $\alpha= 1,...,d$ along the $d$ Cartesian axes.
The shear transformation  is achieved by a a rotation by means of
${\bf \widehat U}={\bf U}\big(\{{\bf e}^{{\rm sp}(\alpha)}\}\big)$
with matrix elements  $\widehat U_{\alpha \beta} = e_{\beta}^{{\rm sp}(\alpha)}$
such that these vectors point along the direction of the Cartesian axes,
\begin{eqnarray}
{\bf \widehat U}{\bf e}^{{\rm sp}(\alpha)}={\mbox {\boldmath$\epsilon$}}^{(\alpha)},\;\;{\bf \widehat U}{\mbox {\boldmath$\xi$}}_{0\pm}^{{\rm sp}(\alpha)}=\xi^{{\rm sp}(\alpha)}_{0\pm}{\mbox {\boldmath$\epsilon$}}^{(\alpha)},
\end{eqnarray}
and by a subsequent rescaling of their lengths by means of a diagonal matrix
${\mbox {\boldmath$\widehat\lambda$}}^{-1/2}$. This
 yields the $d$ transformed correlation vectors
\begin{eqnarray}
\label{transcorrsp}
{\mbox {\boldmath$\widehat\xi$}}_{0\pm}^{(\alpha)}
%{\mbox {\boldmath$\widehat\xi$}}_{0\pm}^{{\rm sp}(\alpha)}
&=&{\mbox {\boldmath$\widehat\lambda$}}^{-1/2}{\bf \widehat U}{\mbox {\boldmath$\xi$}}_{0\pm}^{{\rm sp}(\alpha)}
=\widehat \lambda_{\alpha}^{-1/2}\xi^{{\rm sp}(\alpha)}_{0\pm}{\mbox {\boldmath$\epsilon$}}^{(\alpha)}
\end{eqnarray}
where $\widehat\lambda_{\alpha}$
are the diagonal elements of ${\mbox {\boldmath$\widehat\lambda$}}$.
We note that the orthogonal matrix ${\bf \widehat U}$ contains $d(d-1)/2$ independent matrix elements which are needed to specify the directions ${\bf e}^{{\rm sp}(\alpha)}$ of the principal axes. These $d$ mutually orthogonal axes can be described by $d(d-1)/2$ independent angles $\Omega_i$, $i=1,2,..., d(d-1)/2$, i.e., one angle $\Omega$ in two dimensions, three angles in three dimensions, etc.,
\begin{eqnarray}
\label{Ud}
{\bf U}\big(\{{\bf e}^{{\rm sp}(\alpha)}\}\big)= {\bf U}\big(\Omega_1,\Omega_2, \Omega_3,...\big).
\end{eqnarray}
This statement holds also for the $\varphi^4$ model.
In order to obtain a system that is isotropic in the scaling region it is necessary and sufficient that $\widehat\lambda_{\alpha}^{1/2} $ is proportional to $\xi^{{\rm sp}(\alpha)}_{0\pm}$. Thus we may make the choice \cite{dohm2023}
\begin{eqnarray}
\label{choiceyyxx}
\widehat\lambda_{\alpha}^{1/2} =  \xi^{{\rm sp}(\alpha)}_{0\pm}/\xi^{\rm iso}_{0\pm}
\end{eqnarray}
where $\xi^{\text { iso}}_{0+}$ and $\xi^{\text { iso}}_{0-}$ are free parameters, together with the requirement \cite{dohm2023} that
\begin{eqnarray}
\label{isoratio}
\xi^{\text { iso}}_{0+}/\xi^{\text { iso}}_{0-}
%=\frac{\xi^{\text { iso}}_{0-}}{\xi^{\text { iso}}_{0+}}=\frac{\xi^{(\alpha)}_{0-}}{\xi^{(\alpha)}_{0+}}
=X_\xi={\rm universal}.
\end{eqnarray}
This requirement is necessary in order to comply with the fact implied by two-scale-factor universality that the ratio of the correlation lengths above and below $T_c$ of any isotropic system near $T_c$ must satisfy the universal relation (\ref{ratioxi}). Eqs. (\ref{transcorrsp})-(\ref{isoratio}) transform the $d$ different principal correlation vectors (\ref{corrvectorsp}) to the $d$ vectors
\begin{eqnarray}
\label{transiso}
{\mbox {\boldmath$\widehat\xi$}}_{0\pm}^{(\alpha)}
=\xi^{\text { iso}}_{0\pm}{\mbox {\boldmath$\epsilon$}}^{(\alpha)}, \;\;\;|{\mbox {\boldmath$\widehat\xi$}}_{0\pm}^{(\alpha)}|
=\xi^{\text { iso}}_{0\pm},\;\;\; \alpha=1,...,d,
\end{eqnarray}
with the angular-independent lengths $\xi^{\text { iso}}_{0+}$ and $\xi^{\text { iso}}_{0-}$ above and below $T_c$, respectively, thus representing an isotropic system \cite{dohm2023} .

We apply this generalized shear transformation to the lattice vectors ${\bf x}$ of the anisotropic system
\begin{eqnarray}
\label{trans}
{\bf \widehat x}= {\mbox {\boldmath$\widehat\lambda$}}^{-1/2} {\bf U}{\bf x}.
\end{eqnarray}
This generates an $n$-vector model on a transformed lattice with lattice points ${\bf \widehat x}$ and  with isotropic correlations characterized by the correlation length $\xi^{\text { iso}}_{0\pm}$ for which two-scale-factor universality can be invoked.
Our generalized shear transformation differs significantly from the special transformation (\ref{shearalt})-(\ref{shearu}) in that our transformation is a pure coordinate transformation without transforming the field $\varphi$ and the coupling $u_0$. As a consequence, this transformation leaves the amplitude of the order-parameter correlation function $G^{\rm sp}$ invariant. Thus the special transformation (\ref{isocorrf}) is replaced by the invariance
\begin{eqnarray}
\label{Gplussp}
G^{\rm sp}({\bf x}, t)
=\widehat G(|{\bf \widehat x}|,t)
 =\widehat G(|{\mbox {\boldmath$\widehat\lambda$}}^{-1/2} {\bf  U}{\bf x}|, t)
\end{eqnarray}
where the transformed  correlation function $\widehat G$ has the isotropic structure (\ref{3c}),
\begin{eqnarray}
\label{3cnvector}\widehat G(|{\bf \widehat x}|, t) &=& \frac{\widehat\Gamma_+(\xi^{\rm iso}_{0+})^{-2+\eta}}{ | {\bf \widehat x} |^{ d - 2 +\eta}}
\;\Psi_\pm \Big(\frac{|{\bf \widehat x}|}{ \xi^{\rm iso}_{\pm}(t)}\Big) \; ,\;\;\;\;\;\;
\end{eqnarray}
with $\widehat\Gamma_+\equiv \Gamma^{\rm iso}_+$
and with the universal scaling function $\Psi_\pm$. The susceptibility $ \chi^{\rm sp}(t)= \Gamma_\pm\; |t|^{-\gamma}$ of the anisotropic system, however, is not invariant under the generalized shear transformation
but is transformed as \cite{dohm2023}
\begin{eqnarray}
\label{relationchi}
\widehat \chi(t)&=&\int d^d{\bf \widehat x}\;\;\widehat G ({\bf \widehat x}, t)=\widehat \Gamma_\pm\; |t|^{-\gamma}\\
&=& (\det {\mbox {\boldmath$\widehat\lambda$}})^{-1/2}\int d^d{\bf x}\;\;G^{\rm sp} ({\bf x}, t)
=\big(\xi^{\rm iso}_{0\pm}/\bar \xi^{\rm sp}_{0\pm}\big)^{d}\chi^{\rm sp}(t)\nonumber\\
\end{eqnarray}
where $\bar \xi^{\rm sp}_{0\pm}$ is the amplitude of the mean correlation length
$\bar \xi^{\rm sp}_\pm(t)=\bar \xi^{\rm sp}_{0\pm}|t|^{-\nu}$. This implies the amplitude relation
$\widehat \Gamma_\pm=\Big(\xi^{\rm iso}_{0\pm}/\bar \xi^{\rm sp}_{0\pm}\Big)^{d}\;\Gamma_\pm^{\rm sp}$.
It has been shown \cite{dohm2023} that by inverting the shear transformation one obtains from (\ref{Gplussp}) the scaling form of the anisotropic  bulk correlation function of the $n$-vector model  for $2\leq d < 4$
\begin{eqnarray}
\label{Gneuxxalt}
G^{\rm sp}({\bf x}, t)&=&\frac{\Gamma_+^{\rm sp}\big(\bar \xi^{\rm sp}_{0+}\big)^{-2+\eta}}{ [{\bf x}\cdot \big({\bf \bar A}^{\rm sp}\big)^{-1}{\bf x}]^{(d-2+\eta)/2}} \nonumber \\
& \times & \Psi_\pm \Big(\frac{[{\bf x}\cdot \big({\bf \bar A}^{\rm sp}\big)^{-1}{\bf x}]^{1/2}}{\bar \xi^{\rm sp}_\pm(t)}\Big)\;\;\;\;\;\;\;\;\;
 \end{eqnarray}
with the same universal scaling function $ \Psi_\pm $ as for
isotropic systems in the $(d,n)$ class. The reduced anisotropy matrix of the $n$-vector model is given by
\begin{eqnarray}
\label{Aquerxx}
%{\bf \bar A}(q,\Omega)=
{\bf \bar A}^{\rm sp}&=&{\bf U}^{-1}{\bf \bar{\mbox {\boldmath$\lambda$}}}^{\rm sp} {\bf U}
={\bf \bar A}\big(\{\xi_{0\pm}^{{\rm sp}(\alpha)},{\bf e}^{{\rm sp}(\alpha)}\}\big)\\
\label{Aquerxxyy}
&=&{\bf U(\{{\bf e}^{{\rm sp}(\alpha)}\})}^{-1}{\bf \bar{\mbox {\boldmath$\lambda$}}}^{\rm sp} \big(\{\xi_{0 \pm}^{{\rm sp}(\alpha)}\}\big) {\bf U(\{{\bf e}^{{\rm sp}(\alpha)}\})}\;\;\;\;\;\;
\end{eqnarray}
 with the diagonal reduced rescaling matrix
\begin{eqnarray}
\label{lambdaquerx}
{\mbox {\boldmath$\bar\lambda$}}^{\rm sp}&=&{\mbox {\boldmath$\widehat\lambda$}}/\Big(\det {\mbox {\boldmath$\widehat\lambda$}}\Big)^{1/d}\;.\;
\end{eqnarray}
It has the diagonal elements
\begin{eqnarray}
\label{lambdaalphaxixyy}
&&\bar \lambda^{\rm sp}_\alpha=\prod^d_{\beta=1,\; \beta \neq\alpha}\Big(\frac{\xi_{0\pm}^{{\rm sp}(\alpha)}}{\xi_{0\pm}^{{\rm sp}(\beta)}}\Big)^{2/d}
=\Big(\frac{\xi_{0\pm}^{{\rm sp}(\alpha)}}{\bar \xi^{\rm sp}_{0\pm}}\Big)^2\;\;\;\;
\end{eqnarray}
where the free parameter $\xi^{\text { iso}}_{0\pm}$ has been canceled.
Thus ${\bf \bar A}^{\rm sp}$ has the same structure as the matrix ${\bf \bar A}$ in (\ref{barAxyy}) for the $\varphi^4$ model but with the arguments $\xi_{0 \pm}^{(\alpha)}$ and ${\bf e}^{(\alpha)}$ replaced by $\xi_{0 \pm}^{{\rm sp}(\alpha)}$ and ${\bf e}^{{\rm sp}(\alpha)}$.
 This can be extended to any weakly anisotropic system beyond the $\varphi^4$ theory without explicit knowledge of an anisotropy matrix ${\bf A}$ as a function of the couplings. This proves the validity of multiparameter universality for the anisotropic bulk correlation function for general $d$ and $n$ with $d(d+1)/2+1$ nonuniversal parameters but a universal scaling function $\Psi_\pm$ and a universal structure of the reduced anisotropy matrix ${\bf \bar A}^{\rm sp}$. Here we have not made any assumptions other than the validity of two-scale-factor universality for isotropic bulk systems and the existence of principal correlation lengths and principal axes for weakly anisotropic systems together with (\ref{ratioalpha}). Our derivation cannot, of course, make any prediction about the dependence of the nonuniversal parameters on the couplings $E_{i,j}$ of the $n$-vector model, unlike the dependence on the couplings $K_{i,j}$ within the $\varphi^4$ model \cite{dohm2019,dohm2018}.

We have shown \cite{dohm2023} that macroscopic measurements of the amplitude of specific heat of anisotropic systems can determine the mean correlation length $\bar \xi^{\rm sp}_{0+}$  appearing in the bulk correlation function (\ref{Gneuxxalt}). Together with macroscopic measurements of the amplitude $\Gamma^{\rm sp}_+$ of the susceptibility of anisotropic systems one obtains the two amplitudes determining the overall amplitude of the bulk correlation function (\ref{Gneuxxalt}).
By contrast, the $d(d+1)/2-1$ nonuniversal parameters contained in the reduced anisotropy matrix ${\bf \bar A}^{\rm sp}$ cannot be determined from macroscopic measurements but require scattering experiments that resolve the nonuniversal angular-dependent correlations. Thus weak anisotropy destroys two-scale-factor universality and introduces a considerable intrinsic diversity of correlation functions in the asymptotic scaling regime with additional five nonuniversal parameters in three dimensions that are generically unknown \cite{night1983,dohm2019,dohm2023} for the anisotropic $n$-vector model (\ref{Hspin}) and for real systems in most cases.
\subsection{Exact bulk  correlation functions in two dimensions}
We apply our analysis to the
anisotropic bulk correlation function of the $(d=2,n=1 )$ universality class with the critical exponents $\nu=1, \eta=1/4$.
This includes both anisotropic scalar $\varphi^4$ models and anisotropic Ising models.
In the isotropic case two-scale-factor universality implies that all of these systems have the same universal structure of the correlation function given in (\ref{3c})
\begin{eqnarray}
\label{3calt2d} G^{\rm iso} (|{\bf x}|, t) &=& \frac{\Gamma^{\rm iso}_+(\xi_{0+}^{\rm iso})^{-7/4}}{ | {\bf x} |^{1/4}} \;\Psi_\pm \Big(\frac{|{\bf x}|}{ \xi^{\rm iso}_{\pm}(t)} \Big), \;\\
\label{isocorr2d}
\xi^{\rm iso}_\pm(t)&=&\xi^{\rm iso}_{0\pm}\;|t|^{-1}, \;\;\;\;\;\; \xi^{\rm iso}_{0+}/\xi^{\rm iso}_{0-}=2,
\end{eqnarray}
with two independent nonuniversal parameters $\Gamma^{\rm iso}_+,\xi_{0+}^{\rm iso}$. Exact correlation functions of the two-dimensional Ising model on several "isotropic lattices"  are known
for a long time \cite{WuCoy,Perk2} but the universal part of the scaling function was left undetermined. To determine the universal scaling function $\Psi_\pm $ it suffices to consider the simplest nontrivial example. This is the standard Ising model with equal nearest-neighbor (NN) couplings on a square lattice \cite{WuCoy}.
Recently we have identified \cite{dohm2019} the universal scaling function  $\Psi_\pm(y_\pm)$ of this isotropic Ising model.

In the anisotropic case the two-dimensional $\varphi^4$ or Ising models have two additional nonuniversal parameters: the angle $\Omega$ or $\Omega^{\rm Is}$ describing the orientation of the two principal axes and the ratio $q=\xi^{(1)}_{0\pm}/\xi^{(2)}_{0\pm}$ or $q^{\rm Is}=\xi^{\rm Is(1)}_{0\pm}/\xi^{\rm Is(2)}_{0\pm}$ of the two principal correlation lengths. These parameters enter the rotation and reduced rescaling matrices
\begin{eqnarray}
\label{Urotation}
&&{\bf{ U}}(\Omega)=\left(\begin{array}{ccc}
 \cos\; \Omega& \sin\; \Omega \\
  -\sin\; \Omega &  \cos\; \Omega \\
\end{array}\right),\;\;\;\;
{\mbox {\boldmath$\bar\lambda$}}(q)
%=\left(\begin{array}{ccc}
% \bar\lambda^{\rm Is}_1& 0 \\
% 0 & \;\bar\lambda^{\rm Is}_2 \\
%\end{array}\right)
=\left(\begin{array}{ccc}
 q& 0 \\
 0 & \;q^{-1} \\
\end{array}\right)\;\;\;\;\;\;\;\;\;\;
\end{eqnarray}
which yield the reduced two-dimensional anisotropy matrix
\begin{eqnarray}
\label{Aquer}
{\bf \bar A}(q,\Omega)&=&{\bf{ U}}(\Omega)^{-1}{\mbox {\boldmath$\bar\lambda$}}(q){\bf{ U}}(\Omega)
\\
&=&
\left(\begin{array}{ccc}
 q \;c_\Omega^2+q^{-1}s_\Omega^2 & \;\;\;(q-q^{-1})\;c_\Omega \;s_\Omega\\
(q-q^{-1})\; c_\Omega\; s_\Omega& q \;s_\Omega^2 +q^{-1}\;c_\Omega^2
\end{array}\right)\;\;\;\;\;\;\;\;\;\;\;\;
\end{eqnarray}
of the anisotropic $\varphi^4$ models and the matrix
\begin{eqnarray}
\label{IsingAquer}
{\bf \bar A}^{\rm Is}(q^{\rm Is},\Omega^{\rm Is})= {\bf \bar A}(q^{\rm Is},\Omega^{\rm Is}),
\end{eqnarray}
of the Ising models, respectively, with the abbreviations $c_\Omega\equiv\cos\Omega,s_\Omega\equiv\sin\Omega$. This leads to the exact anisotropic two-dimensional
correlation functions
\begin{eqnarray}
\label{GneuxxaltIsingz}
G({\bf x}, t) &=& \frac{\Gamma_+\;(\bar \xi_{0+})^{-7/4}}{\big [{\bf x}\cdot \big({\bf \bar A}\big)^{-1}{\bf x}\big]^{1/8}}\;
 \Psi_\pm \Big(\frac{[{\bf x}\cdot \big({\bf \bar A}\big)^{-1}{\bf x}]^{1/2}}{\bar \xi_\pm(t)}\Big)\;\;\;\;\;\;\;\;\;
 \end{eqnarray}
and
\begin{eqnarray}
\label{GneuxxaltIsingz}
G^{\rm Is}({\bf x}, t) &=& \frac{\Gamma^{\rm Is}_+\;(\bar \xi^{\rm Is}_{0+})^{-7/4}}{\big [{\bf x}\cdot \big({\bf \bar A}^{\rm Is}\big)^{-1}{\bf x}\big]^{1/8}}\;
 \Psi_\pm \Big(\frac{[{\bf x}\cdot \big({\bf \bar A}^{\rm Is}\big)^{-1}{\bf x}]^{1/2}}{\bar \xi^{\rm Is}_\pm(t)}\Big)\;\;\;\;\;\;\;\;\;
 \end{eqnarray}
for the anisotropic $\varphi^4$ and Ising models, respectively. They satisfy the feature of multiparameter universality in that they have the same universal scaling function $\Psi_\pm$ and the same universal structure of the anisotropy matrix but depend on four independent nonuniversal parameters $\Gamma_+,\bar \xi_{0+},q,\Omega$  and $\Gamma^{\rm Is}_+,\bar \xi^{\rm Is}_{0+},q^{\rm Is},\Omega^{\rm Is}$, respectively, thus violating two-scale-factor universality. We conclude that our result predicts the exact scaling form of the correlation function of all weakly anisotropic systems of the ($d=2,n=1$) universality class \cite{dohm2023}. This implies that the calculation of the correlation function of any two-dimensional weakly anisotropic system of the Ising universality class no longer requires a new calculation of a scaling function and of an anisotropy matrix but can be restricted to the much simpler task of determining four nonuniversal parameters, namely the amplitude of the susceptibility and the specific heat above $T_c$,  the ratio of the two principal correlation lengths, and the angle $\Omega$ determining the orientation (angles $\Omega$ and $\Omega+\pi/2$) of the two principal axes. This constitutes a fundamental simplification in the analytic theory of two-dimensional anisotropic systems as well as in the analysis of numerical or experimental data of anisotropic correlation functions.

A rigorous confirmation for this conclusion comes from the analysis \cite{dohm2019,dohm2023} of the exact correlation function $G^{\rm tr}({\bf x},t)$ \cite{Vaidya1976}
of the fully anisotropic triangular-lattice Ising model with the Hamiltonian
\begin{eqnarray}
\label{Htr}
H^{\rm tr}&=& \sum_{j,k} [-E_1 \sigma_{j,k}  \sigma_{j,k+1}-E_2 \sigma_{j,k}  \sigma_{j+1,k}\nonumber\\
&-&E_3 \sigma_{j,k}  \sigma_{j+1,k+1}],\\
\sigma_{j,k} &=&\pm 1,
\end{eqnarray}
with horizontal, vertical, and diagonal couplings $E_1,E_2,E_3$ on a square lattice. The condition for a ferromagnetic critical point with weak anisotropy is 
%\cite{Houtappel}
%
\begin{eqnarray}
\label{range}
E_1+E_2 > 0, E_1+E_3 > 0, E_2+E_3 > 0.
\end{eqnarray}
The structure of $G^{\rm tr}({\bf x},t)$ was found \cite{dohm2019,dohm2023} to be in exact agreement with (\ref{IsingAquer}) and (\ref{GneuxxaltIsingz}) in the full range (\ref{range}) of the couplings $E_i$ and the exact nonuniversal parameters $\Gamma^{\rm tr}_+,\;\bar \xi^{\rm tr}_{0+}, q^{\rm tr},\Omega^{\rm tr}$ were determined analytically \cite{dohm2019,dohm2023,DKW2023} in this range. It would be worthwhile to verify the validity of multiparameter universality by analyzing the universality properties of the correlation functions of other two-dimensional anisotropic models \cite{Baxter1982,Perk1,Perk2,Perk3,Perk4,CoyWu}.
\subsection{Angular-dependent correlation vector}
We briefly discuss the notion of an angular-dependent correlation vector introduced recently \cite{dohm2023} that is relevant to both bulk and confined systems with arbitrary boundary condtions.
Since such vectors exist in all weakly anisotropic systems including $\varphi^4$ and Ising models we suppress the superscript $"{\rm Is}"$ in the following.
In a two-dimensional  anisotropic system it is of interest to consider a correlation length as a measure of the spatial range of the critical correlations in a certain spatial direction with an angle $\theta$ described by a unit vector
\begin{eqnarray}
 {\bf e}(\theta) = \left(\begin{array}{c}
  % after \\: \hline or \cline{col1-col2} \cline{col3-col4} ...
  \cos \theta \\
  \sin \theta \\
 \end{array}\right)  . \;
\end{eqnarray}
This can be done by introducing polar coordinates
${\bf x} = r\;{\bf e}(\theta)$
for the argument ${\bf x}$ of the anisotropic bulk correlation function $G({\bf x},t)$.
By rewriting the argument of the scaling function $\Psi_\pm$ in (\ref{GneuxxaltIsingz}) as
\begin{eqnarray}
\frac{[{\bf x}\cdot {\bf \bar A}(q,\Omega)^{-1}{\bf x}]^{1/2}}{\bar \xi_\pm(t)}= \frac{r}{\xi_\pm(t,\theta-\Omega,q)}
\end{eqnarray}
we obtain from (\ref{Aquer}) the angular-dependent correlation length \cite{dohm2019}
\begin{eqnarray}
\label{anguldefin}
&&\xi_{\pm}(t,\theta-\Omega,q)=\xi_{0\pm}(\theta-\Omega,q)|t|^{-1},\\
\label{angulthetaz}
&&\xi_{0\pm}(\theta-\Omega,q)=\frac{\bar \xi_{0\pm}}{f(\theta-\Omega,q)},\\
\label{aaa}
&&f(\theta-\Omega,q)=[q\sin^2(\theta-\Omega)+q^{-1}\cos^2(\theta-\Omega)]^{1/2}.\;\;\;\;\;\;\;\;\;\;\;\;
\end{eqnarray}
This yields the representation of the correlation function in polar coordinates
\begin{eqnarray}
\label{GneuxxaltIsingr}
G({\bf x}, t) &=&
 \frac{\Gamma_+}{\big(\bar \xi_{0+}\big)^{2}}\Bigg(\frac{\xi_{0\pm}(\theta-\Omega,q)}{ r }\Bigg)^{1/4}\;\nonumber\\
 & \times &\Psi_\pm \Big( \frac{r}{\xi_\pm(t,\theta-\Omega,q)}\Big).\;\;\;\;\;\;\;\;\;\;
 \end{eqnarray}
The usefulness of the correlation length $\xi_{0\pm}(\theta-\Omega,q)$ is its property of having two extrema  \cite{dohm2019} with respect to $\theta$ at the angles
\begin{eqnarray}
\theta^{(1)}=\Omega, \; \theta^{(2)}=\Omega+\pi/2
\end{eqnarray}
which determine the principal directions, i. e.,
\begin{eqnarray}
\label{corrvectoreins}
\xi_{0\pm}(0,q)&=&\bar \xi_{0\pm}q^{1/2}=\xi_{0\pm}^{(1)} ,\\
\label{corrvectorzwei}
\xi_{0\pm}(\pi/2,q)
&=&\bar \xi_{0\pm}q^{-1/2}=\xi_{0\pm}^{(2)}.
\end{eqnarray}
Here we extend the definition of the amplitude $\xi_{0\pm}(\theta-\Omega,q)$ to a definition of an {\it angular-dependent correlation vector} \cite{dohm2023}
\begin{eqnarray}
\label{corrvectorangult}
{\mbox {\boldmath$\xi$}}_\pm(t,\theta,\Omega,q)&=&{\mbox {\boldmath$\xi$}}_{0\pm}(\theta,\Omega,q)|t|^{-\nu},\\
\label{corrvectorangul}
{\mbox {\boldmath$\xi$}}_{0\pm}(\theta,\Omega,q)&=&\xi_{0\pm}(\theta-\Omega,q)\;{\bf e}(\theta)
\end{eqnarray}
that is oriented in the direction of the angle $\theta$.
This is a generalization of the principal correlation vectors (\ref{corrvectorsp}) which are obtained from (\ref{angulthetaz})-(\ref{corrvectorangul}) for $\theta=\theta^{(1)}$ and $\theta=\theta^{(2)}$.
It can be verified that the generalized shear transformation
of Sec. IV. B
indeed transforms the angular-dependent correlation vector
(\ref{corrvectorangul}) to a vector with a rescaled {\it isotropic} length $\xi^{\rm iso}_{0\pm}$,
\begin{eqnarray}
\label{transangular}
&{\mbox {\boldmath$\widehat\lambda$}}^{-1/2}& {\bf  U}(\Omega)\;{\mbox {\boldmath$\xi$}}_{0\pm}(\theta,\Omega,q)=\xi^{\rm iso}_{0\pm}\;{\bf e}(\theta-\Omega,q),
\end{eqnarray}
\begin{eqnarray}
 \label{2pp1}
 {\bf e}(\theta-\Omega,q)&=&\frac{1}{\big[1+q^2\tan^2(\theta-\Omega)\big]^{1/2}} \left(\begin{array}{c}
  % after \\: \hline or \cline{col1-col2} \cline{col3-col4} ...
  1  \\
  q\tan(\theta-\Omega)
 \end{array}\right),\nonumber \\
 \\
 |{\bf e}(\theta-\Omega,q)|&=&1,
  \end{eqnarray}
with an orientation that depends on the relative angle $\theta-\Omega$.

The same angular-dependent representation can be employed in
the shear transformation applied to any lattice point ${\bf x}$  of the anisotropic system
with the length $r=|{\bf x}|$. For the generalized shear transformation (\ref{trans}) this yields the transformed vector \cite{dohm2023}
\begin{eqnarray}
\label{transangularx}
{\bf \widehat x}=
&{\mbox {\boldmath$\widehat\lambda$}}^{-1/2}& { \bf  U}(\Omega)\;{\bf x}=\frac{\xi^{\rm iso}_{0\pm}}{\xi_{0\pm}(\theta-\Omega,q)}\;|{\bf x}|\;{\bf e}(\theta-\Omega,q)\;\;\;\;\;\;\;\;\;\;\;\;
\end{eqnarray}
where the factor $\xi^{\rm iso}_{0\pm}/\xi_{0\pm}(\theta-\Omega,q)$ describes the amount of rescaling of the length. For the special shear transformation (\ref{shearalt}) of the $\varphi^4$ model the corresponding representation of the transformed vector ${\bf x'}$ reads \cite{dohm2023}
\begin{eqnarray}
\label{transangularxprime}
{\bf x'}=
&{\mbox {\boldmath$\lambda$}}^{-1/2}& {\bf U}(\Omega)\;{\bf x}=\frac{\xi'_{0\pm}}{\xi_{0\pm}(\theta-\Omega,q)}\;|{\bf x}|\;{\bf e}(\theta-\Omega,q).\;\;\;\;\;\;\;\;\;\;\;\;
\end{eqnarray}
The notion of an angular-dependent representation of correlation vectors and
lattice points is applicable to both bulk and confined systems.
This is of relevance to the application of the shear
transformation to the boundaries of finite systems discussed in Sec. V and in Refs. [63-65]. The discussion given above is also applicable to the Gaussian model \cite{dohm2023}.
\renewcommand{\thesection}{\Roman{section}}
\renewcommand{\theequation}{5.\arabic{equation}}
\setcounter{equation}{0}
\section{Multiparameter universality in finite anisotropic ${\bf d=2}$ systems}
In this section we report on recent advances \cite{DW2021,DWKS2021,dohm2023,DKW2023,KWD2023} in finite-size theory of weakly anisotropic systems with periodic BC in two
%and three
dimensions.
\subsection{Exact critical free energy and Casimir amplitude on a rectangle}
Recently \cite{DW2021} exact results have been derived for the critical free energy
${\cal F}_c$
and the ensuing critical Casimir amplitude
$X_c$
of the anisotropic $\varphi^4$ theory at $T_c$ on a finite  rectangle with periodic BC. These results were based on (i) the special shear transformation (\ref{shearalt})-(\ref{shearu}) of the anisotropic $\varphi^4$ model on a rectangle to an isotropic $\varphi^4$ model on a parallelogram, (ii) the exact critical free energy ${\cal F}^{\rm CFT}_c$ at $T_c$ of
conformal field theory  \cite{franc1997} for the isotropic Ising model on a parallelogram, and (iii) the universality of ${\cal F}^{\rm CFT}_c$ according to the Privman-Fisher hypothesis \cite{pri}.
Corresponding predictions were presented \cite{DW2021} for the finite anisotropic
triangular-lattice Ising model \cite{Vaidya1976,dohm2019} (\ref{Htr}) on the basis of the assumption that multiparameter universality \cite{dohm2018} is valid for this model
but no proof was given. A quantitative test was performed \cite{DWKS2021} by high-precision Monte Carlo simulations for a special Ising model on a square with a diagonal anisotropic coupling and remarkable agreement was found with the predicted \cite{DW2021} amplitude ${\cal F}^{\rm Is}_c$.
Here we report on an analytic proof \cite{dohm2023} for the validity of multiparameter universality for weakly anisotropic Ising models with the general Hamiltonian
\begin{eqnarray}
\label{HspinIsing}
H^{\rm Is}& =& - \sum_{i,j} E_{i,j} \sigma_i \sigma_j,\;\;\; \sigma_i=\pm 1.
\end{eqnarray}
The strategy of this proof is (a) to apply of the generalized shear transformation (\ref{trans}) to the boundaries of the anisotropic Ising model on a rectangle which is transformed to an isotropic Ising model on a parallelogram, (b) to exploit the invariance of the critical free energy under this shear transformation \cite{dohm2023}, (c) to employ the exact result of conformal field theory \cite{franc1997} for the critical free energy of the isotropic system, (d) to invert the shear transformation by substituting the transformation formulae.

We assume a finite $L_\parallel \times L_\perp$ rectangle spanned by the vectors ${\bf L}_\parallel=L_\parallel\;(1,0)$  and ${\bf L}_\perp=L_\perp\;(0,1)$ in the horizontal and vertical directions with the aspect ratio
\begin{eqnarray}
\rho_{\rm rec}=L_\perp/L_\parallel.
\end{eqnarray}
We apply our generalized shear transformation (\ref{trans}) directly to the anisotropic Ising model on this rectangle in order to obtain an isotropic Ising model on a parallelogram.
The transformation applied to the boundaries of the rectangle reads
\begin{eqnarray}
\label{LparaLperp}
{\bf L}_{p \parallel}&=&{\mbox {\boldmath$\widehat\lambda$}}^{-1/2} {\bf U}(\Omega^{\rm Is}){\bf L_\parallel},\;\;
{\bf L}_{p \perp}={\mbox {\boldmath$\widehat\lambda$}}^{-1/2} {\bf U}(\Omega^{\rm Is}){\bf L}_\perp,\;\;\;\;\;\;\;\;\;
\end{eqnarray}
where ${\bf U}(\Omega)$ is given by (\ref{Urotation}) and where ${\mbox {\boldmath$\widehat\lambda$}}$ has
 the diagonal elements
$\lambda^{\rm Is}_1=[\xi^{{\rm Is}(1)}_{0\pm}/\xi^{\rm iso}_{0\pm}]^2,
\lambda^{\rm Is}_2=[\xi^{{\rm Is}(2)}_{0\pm}/\xi^{\rm iso}_{0\pm}]^2\;.$
This
generates an isotropic Ising model with the bulk correlation-length amplitude $\xi^{\rm iso}_{0\pm}$ on a parallelogram spanned by the vectors
${\bf L}_{p \parallel}$ and ${\bf L}_{p \perp} $ with the transformed aspect ratio
\begin{eqnarray}
\label{rhoparaaspect}
\rho_{\rm p}=|{\bf L}_{p\perp }|/|{\bf L}_{p \parallel}|.
\end{eqnarray}
The angle $\alpha$ between the vectors ${\bf L}_{p\perp }$ and ${\bf L}_{p \parallel}$ is
determined by
\begin{eqnarray}
\label{alphapara}
\cos \alpha=\frac{{\bf L}_{p \perp}\cdot{\bf L}_{p \parallel}}{|{\bf L}_{p\perp }||{\bf L}_{p \parallel}|}.
\end{eqnarray}
 This yields the transformed aspect ratio \cite{dohm2023}
\begin{eqnarray}
\label{ratioparazz}
\rho_{\rm{p}}(\rho_{\rm rec},q^{\rm{Is}},\Omega^{\rm{Is}})
\label{lengthratioPRL}
&=& \rho_{\rm rec}\; \Bigg[\frac{\tan^2\Omega^{\rm Is}+(q^{\rm Is})^2}{1+(q^{\rm Is})^2\tan^2\Omega^{\rm Is}}\Bigg]^{1/2}\;\;\;
\\
\label{elementratio}
&=&\rho_{\rm rec}\;
\Bigg[\frac{{\bf \bar A}(q^{\rm Is},\Omega^{\rm Is})_{11}}{{\bf \bar A}(q^{\rm Is},\Omega^{\rm Is})_{22}}\Bigg]^{1/2}
\end{eqnarray}
and the angle $\alpha$ determined by
\begin{eqnarray}
\label{alphaising}
\cot \alpha(q^{\rm Is},\Omega^{\rm Is})
\label{alphaalt}
&=&[(q^{\rm Is})^{-1}-q^{\rm Is}]\cos \Omega^{\rm Is} \sin \Omega^{\rm Is}\\
\label{element21}
&=& - {\bf \bar A}(q^{\rm Is},\Omega^{\rm Is})_{12}.
\end{eqnarray}
No specific properties of the weakly anisotropic Ising model were needed in the derivation of (\ref{ratioparazz}) and (\ref{element21}), thus these relations have a universal structure. In Eqs. (7) and (8) of Ref. [38] the same
formulae as
(\ref{elementratio}) and
(\ref{element21}), but with $q^{\rm Is},\Omega^{\rm Is}$ replaced by $q,\Omega$, were first obtained for the $\varphi^4$ model. These formulae were then adopted \cite{DW2021} for the Ising model by the substitution $q \to q^{\rm Is}, \Omega \to \Omega^{\rm Is}$ on the basis of the hypothesis of multiparameter universality. Here we have provided an analytic justification for this substitution directly within the Ising model, without any assumption and without recourse to the $\varphi^4$ model, through  our generalized shear transformation (\ref{trans}) and (\ref{LparaLperp}) for the Ising model. Thus this transformation between an anisotropic and an isotropic Ising model
specifies what was called "effective shear transformation" of the Ising model in Fig. 1 of Ref. [38].

The critical free energy of the transformed isotropic Ising model on the parallelogram is denoted by ${\cal F}^{\rm Is,iso}_c(\alpha,\rho_{\rm p})$ which, according to the Privman-Fisher hypothesis \cite{pri}, is a  universal function of the geometric parameters $\alpha$ and $\rho_{\rm p}$.  The generalized shear transformation is a smooth coordinate transformation which leaves the singular part of the free energy invariant \cite{dohm2023}. Denoting the critical free energy of the anisotropic Ising model on the rectangle by ${\cal F}^{\rm Is}_c$ we have the exact invariance relation
\begin{eqnarray}
\label{invariance}
{\cal F}^{\rm Is}_c={\cal F}^{\rm Is,iso}_c(\alpha,\rho_{\rm p}).
\end{eqnarray}
It is well known \cite{franc1997} that a parallelogram with periodic BC is topologically equivalent to a torus. As pointed out recently \cite{DW2021} the exact result for ${\cal F}^{\rm Is,iso}_c(\alpha,\rho_{\rm p})$  can be taken directly from the known critical free energy ${\cal F}_c^{\rm{CFT}}(\tau) $ of conformal field theory for the isotropic Ising model on a torus\cite{franc1997}
\begin{eqnarray}
\label{calFCFTx}
&&{\cal F}^{\rm Is,iso}_c(\alpha,\rho_{\rm p})={\cal F}_c^{\rm{CFT}}(\tau)= -\ln Z^{\rm{CFT}}(\tau),
\\
\label{taux}
&&\tau(\alpha,\rho_{\rm{p}}) = \mbox{Re}\; \tau + i\; \mbox{Im} \;\tau= \; \tau_0 + i\; \tau_1=\rho_{\rm{p}}\exp(i\; \alpha)\nonumber\\
\end{eqnarray}
which is characterized by the complex torus modular parameter $\tau(\alpha,\rho_{\rm{p}})$. The $\tau$-dependence of ${\cal F}_c^{\rm{CFT}}(\tau) $ is universal and is expressed in terms of Jacobi theta functions \cite{franc1997,DW2021}
$\theta_i(0|\tau)\equiv \theta_i(\tau)$  as
\begin{equation}
\label{ZIsing}
Z^{\rm{CFT}}(\tau)=\big({|\theta_2(\tau)|+|\theta_3(\tau)|+|\theta_4(\tau)|}\big)/\big({2|\eta(\tau)|}\big),
\end{equation}
with
$\eta(\tau)=\Big[\frac{1}{2}\theta_2(\tau)\theta_3(\tau)\theta_4(\tau)\Big]^{1/3}$.
The crucial step is to transfer this exact information  from the isotropic Ising model to the anisotropic Ising model by means of inverting the generalized shear transformation. This is achieved by defining the $(q^{\rm Is} ,\Omega^{\rm Is})$-dependent torus parameter
\begin{equation}
\label{tauqomega}
\tau(\rho_{\rm rec},q^{\rm Is},\Omega^{\rm Is})= \tau\big(\alpha(q^{\rm Is},\Omega^{\rm Is}),\rho_{\rm{p}}(\rho_{\rm rec},q^{\rm Is},\Omega^{\rm Is})\big)
\end{equation}
with $\alpha(q^{\rm Is},\Omega^{\rm Is})$ and $\rho_{\rm{p}}(\rho_{\rm rec},q^{\rm Is},\Omega^{\rm Is})$ given by (\ref{alphaalt}) and (\ref{elementratio}) and by substituting (\ref{tauqomega}) into the isotropic formula (\ref{calFCFTx}). Using (\ref{invariance}) we then obtain the exact result for the critical free energy and the  Casimir amplitude $X^{\rm Is}_c$ of the anisotropic Ising model on the rectangle
as
\begin{eqnarray}
\label{FIS}
{\cal F}^{\rm Is}_c(\rho_{\rm rec},q^{\rm Is},\Omega^{\rm Is})&=&-\ln Z^{\rm{CFT}}\big(\tau(\rho_{\rm rec},q^{\rm Is},\Omega^{\rm Is})\big), \\
\label{FcasCFT}
X^{\rm Is}_c(\rho_{\rm rec},q^{\rm Is},\Omega^{\rm Is})&=& -\rho_{\rm rec}^2 \;\partial {\cal F}^{\rm Is}_c(\rho_{\rm rec},q^{\rm Is},\Omega^{\rm Is})/\partial\rho_{\rm rec}\;\;\;\;\;\;\;\;\;\;\;\;
\end{eqnarray}
with two nonuniversal parameters $q^{\rm Is},\Omega^{\rm Is}$. Going from (\ref{calFCFTx}) to (\ref{FIS}) is equivalent to performing a nonuniversal inverse shear transformation from the isotropic to the anisotropic system. The universal exact structure of (\ref{FIS}) and (\ref{FcasCFT}) holds for all weakly anisotropic systems in the $(d=2,n=1)$ Ising universality class and confirms the predictions \cite{DW2021} obtained on the basis of the hypothesis of multiparameter universality. This completes our proof.

We add the following comments.
The difference between (\ref{calFCFTx}) and (\ref{FIS}) is that the former relation contains a universal function of the geometric variables $\alpha$ and $\rho_{\rm p}$ in agreement with two-scale factor universality \cite{pri} whereas (\ref{FIS}) contains  a universal function  reflecting multiparameter universality with the two  nonuniversal anisotropy parameters $q^{\rm Is}$ and $\Omega^{\rm Is}$.
The same structure was derived \cite{DW2021} for the $\varphi^4$ model without any assumption other than two-scale-factor universality for isotropic systems.
This establishes the universality of the complex self-similar structures of the critical free energy and the Casimir amplitude discovered \cite{DW2021} for all weakly anisotropic systems with periodic BC in this universality class.

As noted already in the context of the anisotropic bulk correlation function,  there exists an intrinsic diversity in finite weakly anisotropic systems as compared to isotropic systems: It arises from the nonuniversal anisotropy parameters $q,\Omega$ and $q^{\rm Is},\Omega^{\rm Is}$ that do not exist in isotropic systems. Furthermore there is a basic difference between $q(\{K_{i,j}\})$ and $\Omega(\{K_{i,j}\})$ of the $\varphi^4$ model (which are known exactly as functions of  $K_{i,j}$  \cite{dohm2019}) and $q^{\rm Is}(\{E_{i,j}\})$ and $\Omega^{\rm Is}(\{E_{i,j}\})$  which are generically unknown \cite{night1983,dohm2019} for the general Ising model (\ref{HspinIsing}).
We conclude that weak anisotropy destroys the universality of the critical Casimir amplitude $X_c$ of isotropic systems and makes $X_c$ to become an unknown quantity for general anisotropic systems whose principal angles and correlation lengths are unknown.
The triangular-lattice Ising model \cite{Vaidya1976,dohm2019} (\ref{Htr}) is a very rare example for a system other than the $\varphi^4$ model for  which these parameters are known exactly as a function of the microscopic couplings \cite{dohm2019,dohm2023}. It would be worthwhile to extend this knowledge to other anisotropic Ising models \cite{Perk1,Perk2,Perk3,Perk4} whose correlation functions have been calculated but whose anisotropy parameters $q^{\rm Is}$ and $\Omega^{\rm Is}$ are as yet unknown.
\subsection{Exact excess free energy  on a rectangle away from $T_c$}
A complete understanding of finite-size properties of weakly anisotropic two-dimensional systems requires to extend the theory to the whole scaling region above and below $T_c$. Exact results for the excess free energy in an anisotropic rectangular geometry away from $T_c$ are available for the Gaussian model \cite{dohm2008,dohm2018} and the Ising model \cite{NO1999} where in the latter case the result was presented in a nonuniversal form. A more general analysis of the universality properties away from $T_c$ has been achieved in forthcoming papers \cite{DKW2023,KWD2023}. Here we report on some of the results of these papers.

We consider the anisotropic Ising model (\ref{HspinIsing}) on a square lattice at $T\neq T_c$ with the shape of the rectangle with the area
\begin{eqnarray}
\label{Vrec}
V_{\rm rec}= L_\parallel L_\perp .
\end{eqnarray}
The exact excess free energy of this model is determined via the same transformation formulae (\ref{ratioparazz}) and (\ref{alphaising}) as in the case $T=T_c$ but in addition one needs a $t$-dependent scaling variable. For $T\neq T_c$ it is the transformation of the area $V_{\rm rec}$ of the rectangle to the area
\begin{eqnarray}
\label{Vpara}
V^{\rm iso}\equiv V_{\rm p}=|{\bf L}_{p \parallel}|\:|{\bf L}_{p\perp }| \;\sin \alpha
\end{eqnarray}
of the isotropic parallelogram that will enter the $t$-dependent scaling variable, as we shall see in (\ref{scalinvariableinvariance}) below. The generalized shear transformation (\ref{trans}) yields the relation between these areas \cite{dohm2023}
\begin{eqnarray}
\label{Vtranssp}
V^{\rm iso} &=&(\det {\mbox {\boldmath$\widehat\lambda$}})^{-1/2}\;V_{\rm rec}=\big(\xi^{\text {iso}}_{0\pm}/\bar \xi^{\rm Is}_{0\pm}\big)^2\;V_{\rm rec}
\end{eqnarray}
where we have used (\ref{choiceyyxx}). This implies the invariance of the area ratios
\begin{eqnarray}
\label{volinvarsp}
\frac{ V^{\rm iso}}{\big(\xi^{\text {iso}}_{0+}\big)^2} &=&\frac{V_{\rm rec}}{\big(\bar \xi^{\rm Is}_{0+}\big)^2}, \;\frac{ V^{\rm iso}}{\big(\xi^{\text {iso}}_{0-}\big)^2} =\frac{V_{\rm rec}}{\big(\bar \xi^{\rm Is}_{0-}\big)^2}
\end{eqnarray}
where
$\big(\xi^{\text {iso}}_{0\pm}\big)^2=V^{\rm iso}_{\rm cor,\pm}$
and
$\big(\bar \xi^{\rm Is}_{0\pm}\big)^2=\xi_{0\pm}^{{\rm Is}(1)}\xi_{0\pm}^{{\rm Is}(2)}=V_{\rm cor\pm}$
are the isotropic (spherical) and anisotropic (ellipsoidal) correlation areas above and below $T_c$, respectively.

For the transformed isotropic system the Privman-Fisher scaling form (\ref{freesing}) for the excess free-energy density $f^{\rm ex,iso} ={\cal F}^{\rm ex, iso}/V^{\rm iso}$ at $h=0$ reads
\begin{eqnarray}
\label{exfreesingisox}
f^{\rm ex,iso} = L_0^{-2} \; F^{\rm ex,iso}\big(C_1 t L_0, \alpha, \rho_{\rm{p}}\big)
\end{eqnarray}
where $F^{\rm ex,iso}$ is a universal function. Here we have set $\nu=1$ for the $d=2$ Ising universality class. The metric factor $C_1$ and the characteristic lengths $L_0$ can be chosen as
\begin{eqnarray}
\label{C1}
C_1 &=& (\xi^{\rm iso}_{0+})^{-1},\;\;
L_0=(V^{\rm iso})^{1/2}.
\end{eqnarray}
Then we obtain
\begin{equation}
f^{\rm ex,iso} =(V^{\rm iso})^{-1}\; F^{\rm ex,iso}\big(\widetilde x(t),\alpha, \rho_{\rm p}\big)
\end{equation}
with the scaling variable
\begin{eqnarray}
 \label{scalinvariablezzy}
 \widetilde x(t)=t\big[V^{\rm iso}/(\xi^{\rm iso}_{0+})^2\big]^{1/2}.
 \end{eqnarray}
This yields the  excess free energy of the isotropic parallelogram
\begin{equation}
\label{universalfexyzz}
{\cal F}^{\rm ex,iso} =V^{\rm iso}\;f^{\rm ex,iso}= F^{\rm ex,iso}\big(\widetilde x(t),\alpha, \rho_{\rm p}\big).
\end{equation}
As discussed above for the case $T=T_c$ the shape dependence can be parameterized in terms of the complex torus modular parameter $\tau(\alpha,\rho_{\rm{p}})$, (\ref{taux}), as given in (\ref{calFCFTx}),
\begin{eqnarray}
\label{Fc}
F^{\rm ex,iso}(0, \alpha, \rho_{\rm p})={\cal F}_c^{\rm{CFT}}\big(\tau(\alpha,\rho_{\rm{p}})\big).
\end{eqnarray}
It is suggestive to conjecture that a description of the shape dependence in terms of $\tau$ is more generally valid for $\widetilde x(t) \neq 0$ such that the excess free energy (\ref{universalfexyzz}) of isotropic systems on a parallelogram  can be written in the scaling form
\begin{eqnarray}
{\cal F}^{\rm ex, iso}={\cal F}^{\rm  ex, iso}(\widetilde x(t),\tau)
\end{eqnarray}
in the whole asymptotic scaling region above and below $T_c$. This is indeed the case as will be substantiated below. We denote the excess free energy  of the anisotropic Ising model (\ref{HspinIsing}) on the rectangle by ${\cal F}^{\rm Is,ex}$. Since the shear transformation does not change the excess free energy \cite{dohm2023} we have the invariance relation
\begin{eqnarray}
\label{spinvar}
%{\cal F}^{\rm iso}_{\varphi\rm{s}}&=&{ \cal F}_{\varphi\rm{s}},\\
\label{Fiso}
{\cal F}^{\rm Is,ex}={\cal F}^{\rm  ex, iso}(\widetilde x(t),\tau)
\end{eqnarray}
where $\widetilde x(t)$ and $\tau$ have to be expressed in terms of the parameters $q^{\rm Is},\Omega^{\rm Is},\rho_{\rm rec},V_{\rm rec}/\big(\bar \xi^{\rm Is}_{0+}\big)^2$ of the anisotropic system. For $\tau$ this has already been done in (\ref{tauqomega}) by substituting the transformation formulae (\ref{ratioparazz}) and (\ref{alphaising}). For the scaling variable  $\widetilde x(t)$ this is achieved by observing its invariance under the shear transformation, namely
\begin{eqnarray}
 \label{scalinvariableinvariance}
 \widetilde x(t)=t\big[V^{\rm iso}/(\xi^{\rm iso}_{0+})^2\big]^{1/2}=t\big[V_{\rm rec}/(\bar \xi^{\rm Is}_{0+})^2\big]^{1/2}
 \end{eqnarray}
as follows from (\ref{scalinvariablezzy}) and (\ref{volinvarsp}). Thus $\widetilde x(t)$ can be written in two ways and can be used as a scaling variable for both the isotropic and anisotropic system. An analogous result was obtained in three dimensions in Eq. (6.12) of Ref. [36].

Finally it is necessary to determine the function ${\cal F}^{\rm  ex, iso}(\widetilde x,\tau)$. According to the Privman-Fisher hypothesis \cite{pri} of two-scale-factor universality, this function is the same for all (isotropic) members of the two-dimensional Ising universality class. Thus it suffices to calculate it for one isotropic system of this universality class which has been performed recently \cite{KWD2023} for the triangular-lattice Ising model (\ref{Htr}).
The exact universal result for isotropic systems in the Ising universality class on a parallelogram in the entire scaling region above and below $T_c$ reads \cite{DKW2023,KWD2023}
\begin{eqnarray}
\label{FEXISO}
&&{\cal F}^{\rm ex, iso}(\widetilde x,\tau)= - \ln \Big\{\frac{1}{2} \big[e^ {{\cal G}(\widetilde x^2,\tau,0,1/2)}+\;e^{{\cal G}(\widetilde x^2,\tau,1/2,0)}
\nonumber\\
&&\;\;\;\;\;\;\;\;\;\;\;\;\;\;\;\;\;\;\;+\;e^{{\cal G}(\widetilde x^2,\tau,1/2,1/2)}-\frac{\widetilde x}{|\widetilde x|} e^{{\cal G}(\widetilde x^2,\tau,0,0)}\big]\Big\}\;\;\;\;\;\;\;\;
\end{eqnarray}
with the function
\begin{eqnarray}
\label{GFEX}
&&{\cal G}(\widetilde x^2,\tau,\Delta_1,\Delta_2)=-\sqrt{\frac{\widetilde x^2 \tau_1}{8\pi^2}} \nonumber\\
&&\times\sum_{m,n=-\infty \atop (m,n)\neq (0,0)}^\infty
 \frac{K_1\left( |m+n\tau| \sqrt{\frac{\widetilde x^2}{ 2\tau_1}}\;\right)}{|m+n\tau|}e^{-2\pi i(m\Delta_1+n\Delta_2)}\;\;\;\;\;\;\;\;\;
\end{eqnarray}
with $ \tau_1 =\mbox{Im} \;\tau >0$ where
\begin{eqnarray}
\label{Bessel}
K_\mu(y)=\frac{1}{2}\Big(\frac{y}{2}\Big)^\mu\int_0^\infty \frac{dz}{z^{\mu+1}}\exp\bigg(-z - \frac{y^2}{4 z}\bigg)
\end{eqnarray}
is the modified Bessel function of the second kind.

The only nonuniversal parameter of this result is the amplitude of the isotropic correlation length above $T_c$ contained in the isotropic scaling variable $\widetilde x$, (\ref{scalinvariablezzy}), with $\nu=1$. This exact information on the isotropic system can be transferred to the anisotropic system by inverting the shear transformation, i.e., by
substituting the transformation formulae (\ref{ratioparazz}), (\ref{alphaising}) and (\ref{scalinvariableinvariance}) into the invariance relation (\ref{spinvar}). Then we arrive at the exact universal scaling form of the excess free energy  of the  anisotropic Ising model (\ref{Htr}) on the rectangle above, at, and below $T_c$ \cite{DKW2023}
\begin{eqnarray}
\label{FISANISOx}
&&{\cal F}^{\rm Is,ex}(\rho_{\rm rec},q^{\rm Is},\Omega^{\rm Is},t V_{\rm rec}^{1/2}/\bar \xi^{\rm Is}_{0+})\nonumber\\
&&={\cal F}^{\rm  ex, iso}\big(t V_{\rm rec}^{1/2}/\bar \xi^{\rm Is}_{0+},\tau(\rho_{\rm rec},q^{\rm Is},\Omega^{\rm Is})\big)
\end{eqnarray}
with three nonuniversal parameters $q^{\rm Is},\Omega^{\rm Is},\bar \xi^{\rm Is}_{0+}$ whereas two-scale-factor universality would allow only one nonuniversal parameter $C_1$ (at $h=0$). No specific properties of the Ising model have been used in the derivation of (\ref{FISANISOx}). Thus this result is universally valid for all systems in the $d=2,n=1$ Ising universality class including the scalar $\varphi^4$ model, in agreement with multiparameter universality, with three different  nonuniversal parameters $q,\Omega,\bar \xi_{0+}$. The latter are known explicitly for the $\varphi^4$ model \cite{dohm2019} as a function of the microscopic couplings $K_{i,j}$ and the lattice structure through the matrix elements (\ref{2i}) of  ${\bf A}$  and the correlation length $\xi'_{0+}(u'_0)$ of the Hamiltonian (\ref{2zxx}). For the general anisotropic Ising model (\ref{HspinIsing}), however, the nonuniversal parameters $q^{\rm Is},\Omega^{\rm Is},\bar \xi^{\rm Is}_{0+}$ are generically unknown and need to be determined for each specific model under consideration. Plotting (\ref{FISANISOx}) in a contour plot in the $q^{\rm Is}-\Omega^{\rm Is}$ plane yields universal self-similar structures \cite{DKW2023} for $t\neq 0$ similar to those found previously \cite{DW2021} for $t= 0$.

For the special case of the Ising model (\ref{Htr}) one can also plot explicitly \cite{DKW2023} the nonuniversal scaling functions ${\cal F}^{\rm{Is},ex}[t,\rho_{\rm rec},E_1/E_3,E_2/E_3]$ and $X^{\rm{Is}}[t,\rho_{\rm rec},E_1/E_3,E_2/E_3]$ for $t\neq 0$ (compare Fig. 5 of Ref. [38] for $t=0$).
These functions are of interest for a comparison with the exact result of Nash and O'Connor \cite{NO1999} for ${\cal F}^{\rm{Is},ex}$ who use a different nonuniversal representation. Exact agreement is found \cite{DKW2023} with their result, and the exact equivalence of (\ref{FEXISO})-(\ref{Bessel}) with the different representation of Ref. [18] is proven analytically in Ref. [65].

The analysis given above can also be extended to the Gaussian model with periodic and antiperiodic BC \cite{DKW2023}.
\subsection{Exact excess free energy  on a parallelogram}
The above results can be extended to anisotropic systems on an arbitrarily shaped parallelogram \cite{DKW2023}. We consider the Ising model (\ref{HspinIsing}) an a parallelogram lattice with the shape of a parallelogram spanned by the vectors
\begin{eqnarray}
\label{vectorspara}
 {\bf {L}_{\rm{1}}} = L_1\;\left(\begin{array}{c}
  % after \\: \hline or \cline{col1-col2} \cline{col3-col4} ...
  1 \\
  0 \\
 \end{array}\right)   \;,\;
  {\bf {L}_{\rm{2}}} = L_2\;\left(\begin{array}{c}
  % after \\: \hline or \cline{col1-col2} \cline{col3-col4} ...
  \cos \vartheta \\
  \sin \vartheta \\
 \end{array}\right)
\end{eqnarray}
in the horizontal direction and in the direction of the angle $\vartheta$, respectively. The aspect ratio $\rho_{\rm{para}}$ and the area $V_{\rm{para}}$ of the parallelogram are given by
\begin{eqnarray}
\label{rhop}
\rho_{\rm{para}}=\frac{|{\bf {L}_{\rm{2}}}|}{|{\bf {L}_{\rm{1}}}|}=\frac{L_2}{L_1},\;\;
\label{area}
 V_{\rm{para}}=L_1 L_2 \sin \vartheta.
\end{eqnarray}
For $\vartheta=\pi/2$ this includes the special case of rectangular geometry at $T_c$ discussed above. The generalized shear transformation applied to the boundaries of the parallelogram reads
\begin{eqnarray}
\label{LparaLperpx}
{\bf L}_{p 1}&=&{\mbox {\boldmath$\widehat\lambda$}}^{-1/2} {\bf U}(\Omega^{\rm Is}){\bf L}_1,\;\;
{\bf L}_{p 2}={\mbox {\boldmath$\widehat\lambda$}}^{-1/2} {\bf U}(\Omega^{\rm Is}){\bf L}_2.\;\;\;\;\;\;\;\;
\end{eqnarray}
This yields a changed isotropic parallelogram which has the transformed area
\begin{eqnarray}
\label{isoareapara}
V^{\rm iso}_{para}&=&|{\bf L}_{p 1}|\;|{\bf L}_{p 2}| \;\sin \alpha_{\rm para}=(\det {\mbox {\boldmath$\widehat\lambda$}})^{-1/2}\;V_{\rm para}\;\;\;\;\;\;\;\;\;\;\;\\
&=&\big(\xi^{\text {iso}}_{0\pm}/\bar \xi^{\rm Is}_{0\pm}\big)^2\;V_{\rm para},
\end{eqnarray}
the transformed aspect ratio
\begin{eqnarray}
\label{Isingratioparastrich}
&&[\widehat\rho_{\rm{para}}(\vartheta,\rho_{\rm{para}},q^{\rm Is},\Omega^{\rm Is})]^2=\frac{|{\bf {L}_{\rm{p 2}}}|^2}{|{\bf {L}_{\rm{p 1}}}|^2}\\
\label{lengthratioPRL}
&&= \rho_{\rm{para}}^2\; \frac{\cos^2(\vartheta-\Omega^{\rm Is})+(q^{\rm Is})^2\sin^2(\vartheta-\Omega^{\rm Is})}{\cos^2\Omega^{\rm Is}+(q^{\rm Is})^2\sin^2\Omega^{\rm Is}},\;\;\;\;\;\;\;\;\;\;
%\\
%\label{elementratio}
%&=&?\rho_{\rm{p}}^2\;\frac{{\bf \bar A}(q^{\rm Is},\Omega^{\rm Is})_{11}}{{\bf \bar %A}(q^{\rm Is},\Omega^{\rm Is})_{22}}?
\end{eqnarray}
and the transformed angle $\alpha_{\rm{para}}$ determined by
\begin{eqnarray}
\label{alphazzz}
&& \cot \alpha_{\rm{para}}(\vartheta,q^{\rm Is},\Omega^{\rm Is})\nonumber\\
&&= \frac{ (q^{\rm Is})^{-1}-q^{\rm Is}\tan \Omega^{\rm Is} \;\tan(\vartheta - \Omega^{\rm Is})}{\tan \Omega^{\rm Is}+ \tan(\vartheta-\Omega^{\rm Is})}.
\end{eqnarray}
The corresponding torus parameter is
\begin{eqnarray}
\label{tauxpara}
\tau(\alpha_{\rm{para}},\widehat\rho_{\rm{para}})& =&  \; \tau_0 + i\; \tau_1=\widehat\rho_{\rm{para}}\exp(i\; \alpha_{\rm{para}})\;\;\;\;\;\\
&=& \tau(\vartheta,\rho_{\rm{para}},q^{\rm Is},\Omega^{\rm Is}),
\label{tauparatheta}
\end{eqnarray}
and the corresponding generalization of the scaling variable (\ref{scalinvariableinvariance}) is
\begin{eqnarray}
 \label{hatscalinvariableinvariance}
 \widehat x(t)=t\big[V^{\rm iso}_{para}/(\xi^{\rm iso}_{0+})^2\big]^{1/2}=t\big[V_{\rm para}/(\bar \xi^{\rm Is}_{0+})^2\big]^{1/2}.
 \end{eqnarray}
Substituting these formulae into the invariance relation (\ref{spinvar}) we obtain the exact universal scaling form of the excess free energy  of the  anisotropic Ising model (\ref{HspinIsing}) on the parallelogram \cite{DKW2023}
\begin{eqnarray}
\label{FISANISO}
&&{\cal F}^{\rm Is,ex}(\vartheta,\rho_{\rm para},q^{\rm Is},\Omega^{\rm Is},t V_{\rm para}^{1/2}/\bar \xi^{\rm Is}_{0+})
\nonumber\\
&&={\cal F}^{\rm  ex, iso}\big(t V_{\rm para}^{1/2}/\bar \xi^{\rm Is}_{0+},\tau(\vartheta,\rho_{\rm para},q^{\rm Is},\Omega^{\rm Is})\big)\;\;\;
\end{eqnarray}
with three nonuniversal parameters $q^{\rm Is},\Omega^{\rm Is},\bar \xi^{\rm Is}_{0+}$  where ${\cal F}^{\rm  ex, iso}(\widehat x, \tau)$ is the same universal function as given in (\ref{FEXISO}). For simplicity we have not changed here the notation for the $\vartheta$-dependent anisotropy parameters $q^{\rm Is},\Omega^{\rm Is},\bar \xi^{\rm Is}_{0+}$  of the Ising model on the anisotropic parallelogram which differ from those on the anisotropic rectangle because the parallelogram-lattice structure of the anisotropy parallelogram at $\vartheta\neq \pi/2$ differs from that of the square-lattice structure of the rectangle. The same comment applies to the nonuniversal parameters $q,\Omega,\bar \xi_{0+}$   of the $\varphi^4$ model. Universal self-similar structures are found \cite{DKW2023} and applications are given to the finite-size part of the specific heat and the Casimir force away from $T_c$. The explicit expressions for all nonunversal parameters of the triangular lattice model (\ref{Htr}) on  a parallelogram lattice \cite{DKW2023} demonstrate a considerable intrinsic diversity caused by weak anisotropy.

In Ref. [64] the analysis given above is extended also to the Gaussian model with periodic and antiperiodic BC.
\renewcommand{\thesection}{\Roman{section}}
\renewcommand{\theequation}{6.\arabic{equation}}
\setcounter{equation}{0}
\section{Field-theoretic approach to ${\bf d=3}$ anisotropic critical behavior:
minimal renormalization without epsilon expansion}
The critical behavior of three-dimensional isotropic and weakly anisotropic systems with short-range interactions is significantly more complex than that of two-dimensional systems, and no exact analytic description is available in three dimensions, except in special cases such as Gaussian models, spherical models, and in the large-$n$ limit.
Here we consider, for finite $n$, the field-theoretic renormalization-group approach
\cite{bre-1,brezin,zinn2007} in the minimal renormalization scheme \cite{hooft-1} at fixed dimension \cite{schl,schl1990,dohm1985,dohm2018}
applied to the $\varphi^4$ model in $2<d<4$ dimensions which is a powerful method to an analytic description of bulk and confined systems near criticality. This approach is applicable to both isotropic and weakly anisotropic systems since weak anisotropy does not generate new ultraviolet divergencies \cite{cd2004,dohm2008}. We restrict ourselves to a brief overview of those aspects that are relevant to finite-size effects in isotropic and weakly anisotropic three-dimensional systems. For an overview of the early history of the  theory of finite-size scaling in isotropic systems see the article by Young in this volume.
\subsection{Lowest-mode separation approach to finite-size theory}
Ordinary perturbation theory for finite systems, in the sense of an expansion around mean-field theory, fails because of unphysical divergencies arising from the isolated lowest ({\bf k}={\bf0}) mode at $T_c$ and from the massless Goldstone modes at the coexistence line below $T_c$. A concept of separating the lowest mode from the higher modes
was formulated within the $4-\varepsilon$ and the $2+\varepsilon$ expansions
\cite{BZ,RGJ}.

We briefly sketch this lowest-mode separation approach based on the isotropic $\varphi^4$ Hamiltonian for general $n$
\begin{eqnarray}
\label{2z}H^{\rm iso}_{\rm field} &=& \int_V d^d x
\big[\frac{r_0} {2} {\bf \varphi}({\bf x})^2 +  \frac{1} {2} (\nabla
{\bf \varphi})^2  +
 u_0 (({\bf \varphi})^2)^2
 %- h' \varphi'
 \big]\;\;\;\;\;\;\;\;
\end{eqnarray}
with $r_0= r_{0c}+ a_0 t$, $a_0>0$,$u_0>0$. Pioneering work on finite-size calculations within this model has been performed previously \cite{BZ,RGJ} for a finite cubic volume $V=L^d$ with periodic BC. In the Fourier expansion
\begin{eqnarray}
\label{fourier}
\varphi({\bf
x}) = L^{-d} \sum_{\bf k} e^{i {\bf k} \cdot {\bf x}_j} \hat
\varphi({\bf k})
\end{eqnarray}
the summations $\sum_{\bf k}$ run over the discrete vectors ${\bf k}$ of the first Brillouin zone of the reciprocal lattice.  It was proposed  to decompose $\varphi({\bf x})$ as
\begin{equation}
\label{decompose}
{\bf \varphi}({\bf x}) = \Phi + \sigma({\bf x})
\end{equation}
into the lowest (homogeneous) mode
\begin{equation}
\label{lowest}
\Phi=L^{-d}\hat\varphi({\bf 0})= L^{-d}\int d^d x {\bf \varphi}
\end{equation}
and the higher inhomogeneous modes
\begin{equation}
\label{higher}
\sigma({\bf x}) =L^{-d} \sum_{\bf k\neq0} e^{i {\bf k} \cdot {\bf x}_j} \hat\varphi({\bf k}).
\end{equation}
Correspondingly the Hamiltonian is decomposed as
\begin{equation}
\label{Hdecomp}
H^{\rm iso}_{\rm field}= H_0(\Phi^2) + \widetilde H (\Phi, \sigma)
\end{equation}
with the zero-mode Hamiltonian
\begin{equation}
\label{zeromodeH}
 H_0 ( \Phi^2) = L^d
\left(\frac{1}{2} r_0 \Phi^2 + u_0 \Phi^4 \right) .
\end{equation}
This yields the partition function
\begin{eqnarray}
\label{4g} Z &=&  \int d^n \Phi \exp \left\{- \left[H_0(\Phi^2) +
{{\Gamma}}(\Phi^2)\right]\right\} \;,\\
\label{4h}
{{\Gamma}} (\Phi^2) &=& -\; \ln \int \; d^n \sigma  \exp \left[ -
\widetilde H (\Phi, \sigma) \right].
\end{eqnarray}
The main task of the theory is the calculation of the contribution of the higher modes arising from ${{\Gamma}}(\Phi^2)$. Some interesting results were obtained within a low-order renormalized treatment of ${{\Gamma}} (\Phi^2)$ based on the $\varepsilon=4-d$ expansion for $T\geq T_c$ \cite{BZ,RGJ} whereas the result for $n=1$ below $T_c$ turned out to be less satisfactory  \cite{EDC}.
This has called for substantial extensions and methodological improvements in order to address several problems of finite-size theory:

(1) an improved perturbation approach capable of determining finite-size scaling functions below $T_c$ not only for $n=1$ \cite{EDC,dohm2008} but also for general $n>1$ including the crossover from the low-temperature Goldstone regime to the critical region \cite{dohm2013,dohm2018},

(2) applications to the order-parameter distribution function above, at, and below $T_c$ for general $n$ \cite{CDS} and in the presence of a finite external field $h$ \cite{cd1998},

(3) applications to the excess free energy and critical Casimir force in cubic \cite{EDC,dohm2008} and block \cite{dohm2011,dohm2018} geometries,

(4) a treatment of a finite-slab geometry with small but finite aspect ratio $\rho << 1$ in order to circumvent the unresolved problem of a dimensional crossover from $d$-dimensional bulk to $d-1$ dimensional film critical behavior \cite{dohm2009,dohm2011},

(5) the extension to realistic Dirichlet BC \cite{dohm2014} that are required to describe the observed critical Casimir force \cite{garcia} near the superfluid transition in isotropic $^4$He films of finite thickness near the minimum below bulk $T_c$,

(6) the prediction of a critical Casimir force scaling function for anisotropic superconducting films not only with idealized periodic BC \cite{wil-1,dohm2018} but also with realistic Dirichlet BC.

It is, of course, beyond the scope of this article to discuss all of these issues. Some general comments are given below and the issue (6)  is further discussed.

\subsection{Minimal renormalization without  $\varepsilon$ expansion \\ in bulk and finite-size theory}
The description of the critical behavior requires perturbative calculations with respect to  $u_0$ as well as multiplicative and additive renormalizations, followed by a mapping of the renormalized quantities from the critical to the noncritical region  where perturbation theory is applicable. These renormalizations are the same in bulk and confined system since there exist no $L$-dependent ultraviolet divergencies \cite{brezin}, and they are the same  in isotropic and in weakly anisotropic systems after an appropriate shear transformation \cite{cd2004,dohm2008}.  The perturbative calculations fall into two categories: those based on the Wilson-Fisher $\varepsilon=4-d$ expansion \cite{wilson}
or the $\varepsilon=d-2$ expansion
\cite{BZ1976}
and those that are carried out at fixed dimensions $d<4$ \cite{zinn2007,dohm1985,schl,schl1990,EDC,dohm2008,dohm2013,dohm2014,dohm2018}. Within these two approaches  one can further distinguish between two types of renormalizations: the use of renormalization conditions \cite{bre-1} and the minimal subtraction scheme \cite{hooft-1}.

The latter renormalization scheme is of particular simplicity and elegance and has the advantage that the renormalization constants (Z-factors) attain the simplest form possible since they absorb just the pole terms $\propto \varepsilon^{-m}$. Furthermore they are the same above and below $T_c$. It has been shown \cite{schl,dohm1985} that the minimally renormalization scheme can be combined with the perturbation theory at fixed dimension (see, e.g., applications  in bulk theory up to four-loop \cite{str2003} and five-loop \cite{larin} order for general $n$ above and below $T_c$ where spurious Goldstone singularities have been canceled). It has been demonstrated \cite{CDS,dohm2013,dohm2018} that this is of particular importance in finite-size calculations below $T_c$ for general $n$ where a conventional $\varepsilon=4-d$ expansion would not be possible because of spurious Goldstone singularities for $n\geq 2$.

We substantiate this point in the context of the lowest-mode separation approach sketched above. The integrand of (\ref{4g}) plays the role of an order-parameter distribution function
\begin{eqnarray}
\label{distribution}
P(\Phi^2)&=&  Z^{-1} \int d^n \sigma \exp \big(- H^{\rm iso}_{\rm field}\big)\\
& = &Z^{-1}\exp \left\{- \left[H_0(\Phi^2) +
{{\Gamma}}(\Phi^2)\right]\right\}
\end{eqnarray}
which is a physical quantity in its own right. The integration $\int d^n \sigma$ in (\ref{distribution}) is to be performed over both longitudinal and transverse fluctuations of $\sigma({\bf x})= \sigma_L({\bf x})+\sigma_T({\bf x})$. Great care must be taken in making approximations in the treatment of the transverse fluctuations $\sigma_T$ in ${{\Gamma}}(\Phi^2)$ such that the exponential structure of $P(\Phi^2)$ is maintained. This means that appropriate approximations \cite{dohm2018} should be made only for the quantity ${{\Gamma}}(\Phi^2)$
without further  expanding $\exp \left[-{{\Gamma}}(\Phi^2)\right]$. This is in line with the minimal subtraction procedure at fixed $d$ \cite{dohm2018} which does not use the fixed point value $u^* \sim O(\varepsilon)$ as smallness parameter. By contrast, the conventional $\varepsilon$ expansion would destroy the exponential structure of $P(\Phi^2)$ and would produce spurious singularities below $T_c$ arising from the transverse fluctuations \cite{dohm2018}. An analytic calculation of $P(\Phi^2)$  for $n=2$ and $n=3$ within the lowest-mode-separation approach combined with the minimal renormalization at fixed dimension $d=3$ in an isotropic cubic geometry has been shown to be in quantitative agreement with Monte Carlo data below $T_c$  \cite{CDS}.
\subsection{Anisotropic extension of the Privman-Fisher scaling form in three dimensions}
We report on the singular part of the free-energy density of the anisotropic $\varphi^4$ model in a finite $L_1\times L_2 \times L_3$ block with periodic BC. By perturbation theory in the minimal renormalization scheme at fixed dimension $d=3$ we have recently derived the anisotropic extension of the Privman-Fisher scaling form (\ref{freesing}) as \cite{dohm2018}
\begin{eqnarray}
\label{freesinganiso}
f^{\rm aniso}_s(t,h,L_1,L_2,L_3) = L_0^{-3} \; Y\Big(\hat x, \hat x_h; \frac{L_1}{L_0},\frac{L_2}{L_0},\frac{L_3}{L_0}, {\bf \bar A}\Big)\nonumber\\
\end{eqnarray}
with the scaling variables
\begin{eqnarray}
\label{xhtilde}
\hat x=t\;(L_0/\bar \xi_{0+})^{1/\nu}\;,
\;\hat x_h= h\;( L_0/\bar \xi_c)^{\Delta/\nu} ,
\end{eqnarray}
and with the reduced anisotropy matrix
\begin{eqnarray}
\label{AquerOmega}
{\bf \bar A} ={\bf \bar A}\big(\{\xi_{0\pm}^{(\alpha)},{\bf e}^{(\alpha)}\}\big)= {\bf \bar A}\Big(\frac{\xi^{(1)}_{0+}}{\bar \xi_{0+}},\frac{\xi^{(2)}_{0+}}{\bar \xi_{0+}}, \frac{\xi^{(3)}_{0+}}{\bar \xi_{0+}},  \Omega_1,\Omega_2,\Omega_3\Big).\nonumber\\
\end{eqnarray}
Here $\bar \xi_{0+}$ and $\bar \xi_c$ are the geometrical mean of the amplitudes $\xi^{(\alpha)}_{0+}$ and $\xi^{(\alpha)}_c$ of the three principal correlation lengths for $T> T_c$, $h= 0$ and for $T=T_c$, $ h\neq 0$ of the anisotropic system. The three angles $\Omega_1, \Omega_2, \Omega_3,$ determine the directions of the  principal unit vectors ${\bf e}^{(\alpha)}, \alpha=1,2,3$ of the three principal axes.
The characteristic length $L_0$ may be chosen as $L_0= V^{1/3}, V=L_1L_2L_3$.
At $h=0$ the anisotropic scaling function $Y$ has been calculated above, at, and below $T_c$ approximately for general finite $n$  and exactly in the large-$n$ limit \cite{dohm2018}.  It describes the entire crossover from the low-temperature to the high-temperature regions including the crossover from Goldstone to the critical behavior.

In the isotropic case we have ${\bf \bar A}={\bf 1}$, and (\ref{freesinganiso})  has the form
\begin{eqnarray}
\label{freesingiso}
f^{\rm iso}_s(t,h,L_1,L_2,L_3) = L_0^{-3} \; Y\Big(\hat x^{\rm iso}, \hat x^{\rm iso}_h; \frac{L_1}{L_0},\frac{L_2}{L_0},\frac{L_3}{L_0}, {\bf  1}\Big)\nonumber\\
\end{eqnarray}
with the scaling variables
\begin{eqnarray}
\label{xhtilde}
\hat x^{\rm iso}=t\;(L_0/\xi^{\rm iso}_{0+})^{1/\nu}\;,
\;\hat x^{\rm iso}_h= h\;( L_0/\xi^{\rm iso}_c)^{\Delta/\nu}
\end{eqnarray}
where $\xi^{\rm iso}_{0+}$ and $\xi^{\rm iso}_c$ are the correlation lengths of the isotropic system.
It has been shown \cite{dohm2018} that $f^{\rm iso}_s(t,0,L_\parallel,L_\parallel,L)$ agrees well with available Monte Carlo data for the isotropic Ising $(n=1)$ and $XY$ $(n=2)$ models in a $L_\parallel^2 \times L$ slab geometry with aspect ratios between $\rho=L/L_\parallel=1/8$ and $\rho=1$.

In the anisotropic case, the reduced anisotropy matrix depends on $\Omega_i$ and on the ratios of  $\xi^{(\alpha)}_{0+}$. Only two of the three ratios $\xi^{(\alpha)}_{0+}/\bar \xi_{0+}$  are independent, thus ${\bf \bar A}$ depends on five independent nonuniversal anisotropy parameters. In addition there are the two mean correlation lengths $\bar \xi_{0+}$ and $\bar \xi_c$ entering the scaling variables $\hat x$ and $\hat x_h$, thus there are seven independent nonuniversal parameters altogether rather than only two parameters in the isotropic case.

Our result \cite{dohm2018} for the scaling function $Y$ is valid for arbitrary Bravais lattices with an arbitrary anisotropy matrix ${\bf A}$ with $\det {\bf A}>0$. It exhibits multiparameter universality within the class of weakly anisotropic three-dimensional $\varphi^4$ models for general $n$. There is essentially no doubt that this result is universally valid also for the $n$-vector model, with seven  nonuniversal parameters, in view of our proof of finite-size multiparameter universality for the two-dimensional Ising universality class and of bulk multiparameter universality for the $(d,n)$ universality classes. Although $Y$  and ${\bf \bar A}$ are  universal functions of their arguments weak anisotropy introduces a substantial degree of nonuniversality through the nonuniversal arguments both in experimental and theoretical analyses. In particular the $d=3$ finite-size amplitude at criticality
\begin{eqnarray}
\label{freesingisoc}
Y_c({\bf \bar A}) =  \; Y\big(0, 0; \rho_1,  \rho_2,  \rho_3, {\bf \bar A}\big),
\end{eqnarray}
the critical Binder cumulant ratio \cite{priv,dohm2008}
\begin{equation}
\label{1f} U_c ({\bf \bar A}) \; =
\frac{1}{3}\;\Big[\frac{\partial^4 Y(0,y;; \rho_1,  \rho_2,  \rho_3,{\bf \bar
A})/ \partial y^4}{(\partial^2 Y(0,y;; \rho_1,  \rho_2,  \rho_3,{\bf \bar A})
/\partial y^2)^2}\Big]_{y=0},
 \;
\end{equation}
and the ensuing critical Casimir amplitude are nonuniversal quantities.
While  $\bar \xi_{0+}$ and $\bar \xi_c$ can be determined by macroscopic thermodynamic measurements of the specific heat and the susceptibility the remaining five nonuniversal parameters $\Omega_i$ and $\xi^{(\alpha)}_{0+}/ \xi^{(\beta)}_{0+}$  require detailed scattering measurements to resolve the angular dependence of the anisotropic critical correlations. On the theoretical side, all seven anisotropy parameters are in principle calculable within the $\varphi^4$ theory from the matrix elements of the anisotropy matrix ${\bf A}$ but this is not easily possible within the $n$-vector model for which no large-distance anisotropy matrix ${\bf A}$ can be defined. The angles $\Omega_i$ of the principal axes are generically unknown functions of the couplings, and their determination in three dimensions for special cases is a nontrivial task as seen already from the research on anisotropic two-dimensional Ising models. Thus, unlike isotropic systems, weakly anisotropic systems exhibit a high degree of intrinsic diversity of critical finite-size behavior within the $(d,n)$ universality classes even if the scaling function $Y$ is known exactly. This affects all observable physical quantities that can be derived from the free-energy density, e.g., the specific heat, the susceptibility, and the critical Casimir force.

So far only a few examples for the anisotropic scaling function \cite{dohm2018} $Y$ and of the ensuing Casimir force scaling function $X$ have been discussed in some detail \cite{dohm2018,DW2021}. It is far beyond the scope of this article to enter a discussion of the complexity of three-dimensional self-similar structures to be expected on the basis of the recent exact analysis of the universal Casimir force scaling function \cite{DW2021} in the case of a planar anisotropy in the three-dimensional anisotropic $\varphi^4$ model with periodic BC in the large-$n$ limit. These complex structures exist not only at $T_c$ but also far below $T_c$ \cite{DW2021} and in the whole scaling region near $T_c$ as demonstrated in a forthcoming paper \cite{DKW2023}.
\subsection{Critical Casimir forces in anisotropic superconductors}
In the past decades there has been much interest in the critical Casimir force in confined systems near criticality which was first predicted by M.E. Fisher and P.G. De Gennes \cite{fisher-gennes} in 1978.
An  experimental verification
has been achieved so far only in isotropic systems, most prominently in superfluid $^4$He films \cite{garcia}.
It has been pointed out by G. Williams \cite{wil-1} that a measurable critical Casimir force should occur also in superconducting films. Superconductors belong to the same universality class as superfluid  $^4$He and have the same (Dirichlet) BC but are anisotropic. Williams' scenario is the following: a superconducting film is connected to a bulk sample of the same material. He argues that below the bulk critical temperature the film-bulk-system can lower its free energy by a transfer of electrons (Cooper pairs) from the film to the bulk system which is analogous to helium atoms moving from the film to the superfluid bulk system. While in the helium system this leads to a thinning of the film the effect in the superconducting film-bulk-system is a transfer of negative electrical charge from the film to the bulk system. Williams argues that this gives rise to an electrical potential difference which can be related to the free-energy difference (per unit area) between the film and the bulk and from which a Casimir force can be derived. He estimates the voltage difference to have a measurable magnitude.

The critical Casimir force is an observable only if the ordering degrees of freedom can enter and leave the system. Therefore it has been claimed in the literature \cite{toldin,diehl2009,diehl2010,DDreview} that this force can be active only in isotropic fluids and that the issue of spatial anisotropy is not relevant in the context of the critical Casimir force. We argue that, unlike the localized degrees of freedom (magnetic moments) of the order parameter of an anisotropic magnetic material, the ordering degrees of freedom (Cooper pairs) of a superconductor are not localized at lattice points but play the role of an electrical superfluid in an anisotropic environment that can leave and enter the film connected to the bulk of the same material, as anticipated by Williams \cite{wil-1}. So far no specific objection has been raised in the literature against this specific argumentation for superconductors, and in a comment \cite{comment} on Ref. [78] the measurability of the critical Casimir force in superconductors has not been questioned. Furthermore, we point to the largely unexplored area of thermodynamic Casimir forces in liquid crystals \cite{singh} which exhibit a wide variety of spatial anisotropy and whose ordering degrees of freedom can leave and enter the system.

In closing we note that experimental studies, Monte Carlo simulations, and further theoretical research are called for in view of the fact that at present no experimental or Monte Carlo data and no analytic predictions are available for the critical Casimir force in anisotropic systems with realistic BC. Also theoretical efforts based on the functional renormalization group \cite{metzner2021} applied to anisotropic confined systems could yield important contributions to this matter. Analytic results for the critical Casimir force in anisotropic films of finite thickness have been presented previously \cite{dohm2018} for the case of periodic BC which refute earlier results \cite{wil-1} where no anisotropy effect in anisotropic superconductors near $T_c$ was found. An analytic renormalization-group study in three dimensions with realistic Dirichlet BC  below $T_c$ without adjustable parameters was performed \cite{dohm2014} that explains the depth and position of the deep minimum of the Casimir force scaling function observed in isotropic $^4$He films \cite{garcia} in the temperature regime  $T_{\rm c,film}<T<T_c$ on a semiquantitative level. It is conceivable that this study can be extended to anisotropic film systems with Dirichlet BC which could lead to quantitative predictions of the critical Casimir force in real superconductors below $T_c$.
\renewcommand{\thesection}{\Roman{section}}
\renewcommand{\theequation}{7.\arabic{equation}}
\setcounter{equation}{0}
\section{Michael Fisher's research and comments on weak anisotropy}
Michael Fisher was interested in lattice structure effects on the critical behavior as seen from his publications
and  from later private communication in spontaneous responses to our research.

In Table IV of the early review article \cite{fish-1} he has listed the weakly anisotropic  interaction types (1) to (4) of short spatial range  with cubic symmetry which exhibit a quadratic, i.e., isotropic momentum dependence at leading order with terms of cubic anisotropy at higher order. This table was later complemented in Ref. [30]. In the work with Tarko \cite{tarko} square, simple-cubic, and body-centered cubic Ising models were studied which demonstrated the universality of several critical bulk amplitude relations that remained unaffected by the cubic weak anisotropy. For the anisotropic Ising model with non-cubic anisotropy Aharony and Fisher \cite{aharony1980} found nonuniversal ratios of correction amplitudes  which in the isotropic case were known to be universal \cite{priv}.
Furthermore, they pointed out \cite{aharony1980} a nonuniversal anisotropy effect on the exact bulk correlation function \cite{CoyWu} of the anisotropic two-dimensional Ising model. This nonuniversality was reanalyzed recently in the framework of multiparameter unversality \cite{dohm2019,dohm2023,DKW2023}.

In 2004 and 2008  Michael Fisher responded spontaneously to our research on weak anisotropy. With regard to the paper \cite{cd2004} by Xiaosong Chen and myself in 2004 he wrote that he felt it "important to have spelled out properly some of the quite subtle implications of lack of cubic symmetry". Then he added a "very mild critical comment" that we have "in some sense overemphasized the degree of nonuniversality". He argued that "there is a reasonable degree of restricted or modified universality in the sense that it is only certain moments of the anisotropy that play a role ", and "the whole issue of universality boils down to how many parameters are needed." Thus, in some sense, he anticipated at an early stage the notion of multiparameter universality with $d(d+1)/2+1$ parameters that was introduced later \cite{dohm2008} and was fully established recently \cite{dohm2018,dohm2019,dohm2023}.

In response to my paper \cite{dohm2008} in 2008 he commented on "your magnum opus on universality"  and  sovereignly admitted the "limitations of universality (especially!) in finite system relative to overoptimistic expectations." He added the "remark that it is the appreciation of the subtleties exhibited by your Figs. 1 and 2 that, many years ago, led David Jasnow and me to hesitate to give a mathematical definition of scaling that one might propose as a 'theorem' suitable for a proof!"

The last time I met Michael Fisher was on a conference in  G\"ottingen 2014 on the occasion of the 80th birthday of Guenter Ahlers. In a conversation with Michael Fisher (together with Pierre Hohenberg) he pointed to the large variety of crystal structures causing nonuniversal anisotropy effects near criticality, and Pierre Hohenberg completely agreed with our analysis of the consequences of weak anisotropy.

\section*{ACKNOWLEDGMENT}
I thank F. Kischel and S. Wessel for useful discussions and collaboration on Refs. [64] and [65].
\end{document}